\documentclass[aps,physrev,reprint,superscriptaddress]{revtex4-2}

\usepackage{graphicx}
\usepackage[caption=false,subrefformat=parens]{subfig}
\usepackage{dcolumn}
\usepackage{bm}
\usepackage{amsmath,amssymb}
\usepackage{mathtools}
\usepackage[colorlinks=true,citecolor=blue,linkcolor=blue,urlcolor=blue]{hyperref}
\usepackage{braket}
\usepackage{longtable}
\usepackage[T1]{fontenc}
\usepackage{color}

\DeclareMathOperator{\Real}{Re}
\DeclareMathOperator{\Imag}{Im}
\DeclareMathOperator{\sgn}{sgn}

\makeatletter
\newcommand\exchange[2]{\let\@tempa#1\let#1#2\let#2\@tempa}
\makeatother

\exchange{\Gamma}{\varGamma}
\exchange{\Delta}{\varDelta}
\exchange{\epsilon}{\varepsilon}
\exchange{\Theta}{\varTheta}
\exchange{\Lambda}{\varLambda}
\exchange{\Xi}{\varXi}
\exchange{\Pi}{\varPi}
\exchange{\Phi}{\varPhi}
\exchange{\Psi}{\varPsi}
\exchange{\Omega}{\varOmega}

\newcommand{\ii}{i}

\begin{document}

\title{Point-gap topology of damped magnon excitations in skyrmion strings}

\author{Yusuke Koyama}
\affiliation{Department of Applied Physics, Nagoya University, Nagoya 464-8603, Japan}
\author{Yuki Kawaguchi}
\affiliation{Department of Applied Physics, Nagoya University, Nagoya 464-8603, Japan}
\affiliation{Research Center for Crystalline Materials Engineering, Nagoya University, Nagoya 464-8603, Japan}

\date{\today}

\begin{abstract}
We theoretically study the non-Hermitian topology of magnons with finite lifetimes due to Gilbert damping.
By incorporating the spin-wave theory and perturbation theory for the Landau-Lifshitz-Gilbert equation including nonlocal damping terms, we analytically evaluate the spectral winding number for point gaps, which indicates the existence of the non-Hermitian skin effect (NHSE).
We find that, in noncollinear magnetic textures, the NHSE can occur even in the absence of nonlocal damping, due to the momentum-dependent mixing of the particle and hole components of magnon eigenmodes, which is absent in collinear magnets with a single magnon band.
In the presence of nonlocal damping along one direction, we show that the winding number for an energy band with a unique minimum is determined from the sign of the wave number at the band minimum.
We demonstrate these results using a model that hosts a skyrmion-string lattice as a steady state.
We further investigate spin-wave propagation dynamics excited by a magnetic-field pulse and show that the propagation direction changes drastically from band to band depending on the presence of local and nonlocal damping, consistent with the nontrivial winding numbers.
\end{abstract}

\maketitle

\section{\label{sec:introduction}Introduction}
Recently, non-Hermitian physics has attracted growing attention~\cite{ashidaNonHermitianPhysics2020,bergholtzExceptionalTopologyNonHermitian2021,okumaNonHermitianTopologicalPhenomena2023,gongTopologicalPhasesNonHermitian2018,kawabataSymmetryTopologyNonHermitian2019,okumaTopologicalOriginNonHermitian2020,zhangCorrespondenceWindingNumbers2020}.
To account for dissipation to environments, effective Hamiltonians governing time evolution of physical systems naturally become non-Hermitian.
Non-Hermiticity of Hamiltonians allows complex spectra, leading to gap structures in the complex-energy plane that are unique to non-Hermitian systems, known as point gaps~\cite{kawabataSymmetryTopologyNonHermitian2019}.
Nontrivial point-gap topology can give rise to the non-Hermitian skin effect (NHSE), in which the spectrum becomes extremely sensitive to boundary conditions and a macroscopic number of eigenstates are localized at the boundaries~\cite{yaoEdgeStatesTopological2018,okumaTopologicalOriginNonHermitian2020,zhangCorrespondenceWindingNumbers2020}.
The NHSE has been experimentally observed in various physical platforms, including photonic~\cite{xiaoNonHermitianBulkBoundary2020,weidemannTopologicalFunnelingLight2020}, mechanical~\cite{brandenbourgerNonreciprocalRoboticMetamaterials2019,ghatakObservationNonHermitianTopology2020,chenRealizationActiveMetamaterials2021}, acoustic~\cite{zhangAcousticNonHermitianSkin2021}, electrical-circuit~\cite{helbigGeneralizedBulkBoundary2020}, and cold-atom systems~\cite{liangDynamicSignaturesNonHermitian2022}.

Such non-Hermitian situations are also encountered in magnons~\cite{shindouTopologicalChiralMagnonic2013,hurstNonHermitianPhysicsMagnetic2022,yuNonHermitianTopologicalMagnonics2024,liNonHermitianLiouvillianSkin2025}.
In the absence of dissipation, the time evolution of free magnons is described by a bosonic Bogoliubov-de Gennes (BdG) Hamiltonian, which is intrinsically not Hermitian but pseudo-Hermitian so as to satisfy the Bose statistics~\cite{shindouTopologicalChiralMagnonic2013,kondoNonHermiticityTopologicalInvariants2020}.
Owing to this property, bosonic systems, including magnons, provide a unique platform for studying non-Hermitian topological phenomena even in the absence of gain or loss.
Nevertheless, it is still important to consider the effect of inevitable damping in magnon systems.
The presence of damping generally breaks the Bose statistics of magnons and thereby the pseudo-Hermiticity of the BdG Hamiltonian as well~\cite{yuNonHermitianTopologicalMagnonics2024,harmsAntimagnonics2024,wangConnectionSpinwavePolarization2025}.
The interplay between intrinsic bosonic properties and extrinsic damping effects may enrich the non-Hermitian topological phases of damped magnons.

To describe spin dynamics in the presence of a realistic damping effect, the Landau-Lifshitz-Gilbert (LLG) equation is widely used~\cite{gilbertPhenomenologicalTheoryDamping2004}.
Although the original LLG equation contains only local Gilbert damping, it can be generalized to include nonlocal damping terms, which have been theoretically studied~\cite{umetsuTheoreticalStudyGilbert2012,thonigGilbertDampingTensor2014,thonigNonlocalGilbertDamping2018,luInfluenceNonlocalDamping2023,reyes-osorioNonlocalDampingSpin2024}.
We note that such nonlocal Gilbert damping is distinct from the dissipative spin-spin coupling mediated by a shared environment~\cite{tserkovnyakExceptionalPointsDissipatively2020,zouBellstateGenerationSpin2022}.
The latter takes the form of nonlocal Gilbert damping only for small-angle dynamics around a collinear equilibrium~\cite{tserkovnyakNonlocalMagnetizationDynamics2005,tserkovnyakExceptionalPointsDissipatively2020}, which is not the case for the noncollinear textures considered in this paper.
A previous study of the NHSE in magnetic systems employed a phenomenological approach based on finite-lifetime broadening in the energy spectrum instead of intrinsic Gilbert damping~\cite{dengNonHermitianSkinEffect2022}.
Another work investigated the NHSE in ferromagnetic multilayer systems described by the LLG equation with nonlocal damping terms~\cite{liNonHermitianLiouvillianSkin2025}.
The analysis in Ref.~\cite{liNonHermitianLiouvillianSkin2025} for a specific ferromagnetic multilayer model implies that the nonlocal damping is essential for the occurrence of the NHSE.
However, the previous studies have focused only on the field-polarized states, and systematic calculations on the winding numbers have yet to be performed.
Accordingly, the theoretical understanding of the NHSE of magnons for various spin states remains incomplete.

In this paper, we elucidate the mechanism of the NHSE of magnons subjected to Gilbert damping within the linear spin-wave theory of the LLG equation including nonlocal damping.
Using the nondegenerate perturbation theory, we show that the presence of damping gives rise to an imaginary part in the magnon eigenenergy.
Notably, an effective damping defined from the imaginary part may differ from the original Gilbert damping, depending on the magnon eigenstates.
This result has already been derived in the presence of only on-site Gilbert damping terms, e.g., in Ref.~\cite{wangConnectionSpinwavePolarization2025}, whereas our derivation extends to the case where nonlocal damping terms are also present.
When the effective damping depends on the wave number $\bm{k}$, the complex spectrum can form a loop in the complex-energy plane, which indicates the occurrence of the NHSE.
Under certain conditions, the effective damping in the momentum space consists of two factors: the Gilbert damping $\alpha_{\bm{k}}$ and $\eta_{n\bm{k}}$, the latter is determined from the relative weights of the particle and hole components of the $n$th magnon eigenstate.
Thus, the NHSE can occur due to not only the presence of nonlocal damping, but also the nontrivial $\bm{k}$ dependence of $\eta_{n\bm{k}}$ even in the absence of nonlocal damping.
The NHSE induced by on-site dissipation itself has been discussed in the literature~\cite{songNonHermitianSkinEffect2019,yiNonHermitianSkinModes2020,okumaQuantumAnomalyNonHermitian2021}.
In our case, the momentum dependence of $\eta_{n\bm{k}}$ originates from the underlying magnetic texture: In noncollinear spin configurations such as skyrmion strings, the magnon spectrum consists of multiple bands, whose eigenmodes mix the particle and hole components depending on the band and the momentum.
Such mixing does not arise in collinear ferromagnets with a single magnon band.
We explicitly discuss this observation by deriving a formula for spectral winding numbers based on the perturbatively obtained expression of the magnon eigenenergy.
In the presence of nonlocal damping, we show that the winding number for each band is determined from the sign of the band minimum point, which also provides an intuitive explanation of the nontrivial spectral windings in Ref.~\cite{liNonHermitianLiouvillianSkin2025}.
We numerically demonstrate these analytical results based on a concrete classical spin model that hosts a skyrmion-string lattice~\cite{sekiPropagationDynamicsSpin2020} as a steady state.
We see that the nontrivial winding numbers essentially determine the damped propagation behavior of the spin waves excited by a magnetic-field pulse.

The rest of this paper is organized as follows.
In Sec.~\ref{sec:formalism}, we derive the non-Hermitian BdG Hamiltonian for the LLG equations with nonlocal damping by using linear spin-wave theory.
In Sec.~\ref{sec:winding_formula}, using nondegenerate perturbation theory, we obtain expressions for the complex eigenvalues and derive a formula for the winding number.
In Sec.~\ref{sec:result_magnon}, based on a classical spin lattice model that hosts a skyrmion-string lattice as a steady state, we validate the winding-number formula and investigate the relationship between the NHSE and spin-wave propagation.
Finally, in Sec.~\ref{sec:conclusion}, we conclude this work.

\section{\label{sec:formalism}Linear spin-wave theory of the LLG equation}
In this section, we develop a spin-wave theory based on the LLG equations with nonlocal damping and derive the corresponding non-Hermitian BdG Hamiltonians.
We use perturbation theory to evaluate the effects of damping on the magnon spectrum.

\subsection{Non-Hermitian Bogoliubov--de Gennes Hamiltonian from the linearized LLG equation}
Throughout this paper, we consider localized spins on a lattice at zero temperature.
Each spin at site $\bm{r}=(x, y, z)$ is assumed to be a classical vector with magnitude $S$, denoted by $\bm{S}_{\bm{r}}\in\mathbb{R}^3$, and the Hamiltonian functional is written as $\mathcal{H}=\mathcal{H}[\bm{S}_{\bm{r}}]$.
The time evolution of each spin $\bm{S}_{\bm{r}}$ is described by a generalized form of the LLG equation,
\begin{align}
    \frac{\partial \bm{S}_{\bm{r}}}{\partial t} &= \gamma \bm{S}_{\bm{r}} \times \bm{B}_{\mathrm{eff}, \bm{r}} - \frac{1}{S} \bm{S}_{\bm{r}} \times \sum_{\bm{r}'} \alpha_{\bm{r}, \bm{r}'} \frac{\partial \bm{S}_{\bm{r}'}}{\partial t}, \label{eq:generalized_LLG}
\end{align}
where $\gamma$ is the gyromagnetic ratio ($\gamma<0$ for electrons).
The first term on the right-hand side of Eq.~(\ref{eq:generalized_LLG}) represents coherent precession around the effective magnetic field defined by $\bm{B}_{\mathrm{eff}, \bm{r}} = -(\hbar\gamma)^{-1} \delta \mathcal{H}[\bm{S}_{\bm{r}}]/\delta \bm{S}_{\bm{r}}$.
The second term accounts for spin relaxation, which can be derived from the Rayleigh dissipation functional given by
\begin{align}
    \mathcal{R}[\bm{S}_{\bm{r}}] &= \frac{\hbar}{2S} \sum_{\bm{r}, \bm{r}'} \frac{\partial \bm{S}_{\bm{r}}}{\partial t} \cdot \alpha_{\bm{r}, \bm{r}'} \frac{\partial \bm{S}_{\bm{r}'}}{\partial t}, \label{eq:dissipation_functional}
\end{align}
where $\alpha_{\bm{r}, \bm{r}'} \in \mathbb{R}$ denotes a phenomenological nonlocal Gilbert damping coefficient between $\bm{r}$ and $\bm{r}'$.

More generally, $\alpha_{\bm{r},\bm{r}'}$ is a $3\times 3$ tensor in spin space that can be determined only microscopically and generally depends even on the spin configuration itself~\cite{brinkerGeneralizationLandauLifshitz2022}.
However, here we treat it as a scalar independent of the spin orientation, as the simplest model of the damping, since our aim is to clarify the effects of damping on magnon excitations rather than to describe a specific material quantitatively.
The matrix $\alpha \coloneqq (\alpha_{\bm{r},\bm{r}'})$, whose elements are labeled by sites, is not necessarily symmetric: Its nonlocal components may contain an antisymmetric part~\cite{brinkerGeneralizationLandauLifshitz2022}.
The antisymmetric part, however, does not contribute to the Rayleigh dissipation functional~(\ref{eq:dissipation_functional}) and hence causes no energy dissipation.
Since our focus is on the effects of dissipation, we hereafter assume $\alpha$ to be real symmetric, $\alpha_{\bm{r},\bm{r}'}=\alpha_{\bm{r}',\bm{r}}\in\mathbb{R}$.
For Eq.~(\ref{eq:generalized_LLG}) to describe the energy-decreasing dynamics, $\alpha$ must be positive semidefinite, denoted by $\alpha\succeq 0$.
When the damping is local, $\alpha_{\bm{r}, \bm{r}'}=\alpha_{\bm{r}} \delta_{\bm{r},\bm{r}'}$, Eq.~(\ref{eq:generalized_LLG}) reduces to the conventional LLG equation~\cite{gilbertPhenomenologicalTheoryDamping2004}.
The case of general $\alpha_{\bm{r}, \bm{r}'}$ has been theoretically studied in the literatures~\cite{thonigNonlocalGilbertDamping2018,luInfluenceNonlocalDamping2023,reyes-osorioNonlocalDampingSpin2024,thonigGilbertDampingTensor2014,umetsuTheoreticalStudyGilbert2012}.
We note that the nonlocal damping considered here is the intrinsic one in the generalized LLG equation, which microscopically originates, e.g., from the nonlocal response of conduction electrons~\cite{thonigNonlocalGilbertDamping2018,luInfluenceNonlocalDamping2023,reyes-osorioNonlocalDampingSpin2024,thonigGilbertDampingTensor2014,umetsuTheoreticalStudyGilbert2012}.
The dissipative coupling between spins mediated by a shared environment, such as spin pumping into a common reservoir~\cite{tserkovnyakExceptionalPointsDissipatively2020,zouBellstateGenerationSpin2022}, is physically distinct and beyond the scope of this work.
Hereafter, we take Eq.~(\ref{eq:generalized_LLG}) as our starting point and will discuss the consequence in terms of non-Hermitian topology in Secs.~\ref{sec:winding_formula} and \ref{sec:result_magnon}.

For convenience, we introduce a local coordinate frame $\{\tilde{\bm{e}}_{\bm{r}}^x, \tilde{\bm{e}}_{\bm{r}}^y, \tilde{\bm{e}}_{\bm{r}}^z\}$ at site $\bm{r}$ such that the $z$ axis is aligned with the direction of the stationary solution $\bm{S}_{\bm{r}}^0$ of Eq.~(\ref{eq:generalized_LLG}), i.e., $\tilde{\bm{e}}_{\bm{r}}^z = \bm{S}_{\bm{r}}^0 / S$.
Spin fluctuations around $\bm{S}_{\bm{r}}^0$ can be parametrized by a complex scalar field $\psi_{\bm{r}}$ as
\begin{align}
    \bm{S}_{\bm{r}} &= \sqrt{S} (\psi_{\bm{r}} \tilde{\bm{e}}_{\bm{r}}^- + \psi_{\bm{r}}^* \tilde{\bm{e}}_{\bm{r}}^+ + \sqrt{S - 2|\psi_{\bm{r}}|^2} \tilde{\bm{e}}_{\bm{r}}^z), \label{eq:spin_wave_param}
\end{align}
where $\tilde{\bm{e}}_{\bm{r}}^\pm = (\tilde{\bm{e}}_{\bm{r}}^x \pm \ii \tilde{\bm{e}}_{\bm{r}}^y)/\sqrt{2}$.
By linearizing Eq.~(\ref{eq:generalized_LLG}) with respect to $\psi_{\bm{r}}$, we obtain the following non-Hermitian Bogoliubov--de Gennes (BdG) equation (see Appendix~\ref{sec:BdG_derivation} for details):
\begin{align}
    \ii\hbar \frac{\partial}{\partial t} \begin{pmatrix}
        \bm{\psi} \\
        \bm{\psi}^*
    \end{pmatrix}
    &= \tilde{H}(\alpha) \begin{pmatrix}
        \bm{\psi} \\
        \bm{\psi}^*
    \end{pmatrix}, \label{eq:BdG}
\end{align}
where $\bm{\psi}\coloneqq (\psi_{\bm{r}})$ denotes a vector whose components are labeled by the site index $\bm{r}$ and
\begin{align}
    \tilde{H}(\alpha) &\coloneqq (\tau_z + \ii A)^{-1} H = (1 + \ii \tau_z A)^{-1} \tilde{H}(0) \label{eq:magnon_nh_hamiltonian}
\end{align}
is the non-Hermitian BdG Hamiltonian with $\tau_{x,y,z}$ being the Pauli matrices acting on the Bogoliubov quasiparticle space.
Here, $H$ and $A$ in Eq.~(\ref{eq:magnon_nh_hamiltonian}) are defined as follows.
$H=H^\dagger$ is the Hermitian Hessian matrix of the spin Hamiltonian $\mathcal{H}$ evaluated at $\bm{S}_{\bm{r}}^0$,
\begin{align}
    H &= \begin{pmatrix}
        h & \Delta \\
        \Delta^* & h^*
    \end{pmatrix}, \label{eq:BdG_hamiltonian}
\end{align}
where
\begin{align}
    h_{\bm{r},\bm{r}'} &= \frac{\delta^2 \mathcal{H}}{\delta{\psi_{\bm{r}'}} \delta{\psi_{\bm{r}}^*}}[\bm{S}_{\bm{r}}^0],\quad \Delta_{\bm{r},\bm{r}'} = \frac{\delta^2 \mathcal{H}}{\delta{\psi_{\bm{r}'}^*} \delta{\psi_{\bm{r}}^*}}[\bm{S}_{\bm{r}}^0].
\end{align}
The Hessian matrix governs the undamped magnon dynamics through $\tilde{H}(0)=\tau_z H$.
$A(\alpha)=A^\dagger(\alpha)$ depends linearly on $\alpha$ and is defined as
\begin{align}
    A_{\bm{r},\bm{r}'} &= \begin{pmatrix}
        \tilde{\bm{e}}_{\bm{r}}^+{}^\top \\
        \tilde{\bm{e}}_{\bm{r}}^-{}^\top
    \end{pmatrix}
    \alpha_{\bm{r},\bm{r}'}
    \begin{pmatrix}
        \tilde{\bm{e}}_{\bm{r}'}^- & \tilde{\bm{e}}_{\bm{r}'}^+
    \end{pmatrix}. \label{eq:A_def}
\end{align}
Here, $A\succeq 0$ follows from $\alpha\succeq 0$.
We also note that the Hermiticity of $A$ follows from the symmetry of $\alpha$.
If $\alpha$ contained an antisymmetric part, Eq.~(\ref{eq:A_def}) would yield an anti-Hermitian contribution to $A$, i.e., a Hermitian contribution to $\ii A$ in Eq.~(\ref{eq:magnon_nh_hamiltonian}).
Such a contribution only shifts the real parts of the magnon energies, consistent with the fact that the antisymmetric part causes no dissipation.

We note that the BdG Hamiltonian (\ref{eq:magnon_nh_hamiltonian}) without damping, $\tilde{H}(0)$, is already non-Hermitian and can have complex eigenvalues.
In the presence of damping, the spectrum of $\tilde{H}(\alpha)$ becomes generally complex even when all eigenvalues of $\tilde{H}(0)$ are real, and their imaginary parts are determined by $\alpha_{\bm{r},\bm{r}'}$, as shown in Eq.~(\ref{eq:perturbed_energy}), obtained using perturbation theory.

Before proceeding to the perturbative calculation of eigenvalues, we discuss general spectral properties of Eq.~(\ref{eq:magnon_nh_hamiltonian}) from a symmetry perspective.
For any $\alpha$ by construction, $\tilde{H}(\alpha)$ possesses anomalous particle-hole symmetry (PHS$^\dagger$), $C' \tilde{H}(\alpha) C'^{-1} = -\tilde{H}(\alpha)$ with $C'=\tau_x K$, where $K$ is the complex-conjugate operator~\cite{kawabataSymmetryTopologyNonHermitian2019}.
It follows that the spectrum of $\tilde{H}(\alpha)$ is symmetric with respect to the imaginary axis; hence, we restrict our attention to eigenenergies with positive real parts (see also Sec.~\ref{subsec:particle_hole}).

In the absence of damping, $\tilde{H}(0)$ additionally possesses pseudo-Hermiticity, $\eta \tilde{H}(0) \eta^{-1} = \tilde{H}^\dagger(0)$ with $\eta\coloneqq \tau_z$, since $H$ is Hermitian~\cite{kawabataSymmetryTopologyNonHermitian2019}.
It follows that the spectrum of $\tilde{H}(0)$ is symmetric with respect to the real axis, i.e, the eigenvalues of $\tilde{H}(0)$ are real or come in complex-conjugate pairs.
Therefore, if a complex eigenvalue appears in the spectrum of $\tilde{H}(0)$, there necessarily exists one with a positive imaginary part, and the corresponding mode grows under an infinitesimal perturbation.
This implies that the initially prepared stationary state is dynamically unstable.
However, in the presence of damping, the pseudo-Hermiticity is generally broken, and thus the complex spectrum of $\tilde{H}(\alpha)$ does not necessarily mean the instability.

\subsection{\label{subsec:particle_hole}Particle and hole bands}
Here, we remark on the notion of the particle and hole bands in the BdG systems.
The energy bands inherently possess a trivial redundancy due to the PHS$^\dagger$ of $\tilde{H}(\alpha)$.
Namely, for each eigenstate $\ket{\phi_n^\mathrm{R}}$ of $\tilde{H}(\alpha)$ with eigenenergy $E_n$, its particle-hole-conjugated counterpart $C' \ket{\phi_n^\mathrm{R}}$ is an eigenstate with eigenenergy $-E_n^*$.
Thus, it suffices to know either $\ket{\phi_n^\mathrm{R}}$ or $C' \ket{\phi_n^\mathrm{R}}$ for eigenvalues with nonvanishing real part.
In this paper, we call an eigenmode $\ket{\phi_n^\mathrm{R}}$ a particle (hole) mode if the indefinite inner product $\braket{\phi_n^\mathrm{R}|\tau_z|\phi_n^\mathrm{R}}$ is positive (negative).
This definition can be justified in terms of the spin-wave excitation energy from stationary states in the absence of damping as follows.
The excitation energy for the spin fluctuation associated with $\bm{\psi}$ is written as $\Delta E = \frac{1}{2} \braket{\psi|H|\psi}$ in the lowest order of $\bm{\psi}$, where $\ket{\psi}\coloneqq \begin{pmatrix}
    \bm{\psi} & \bm{\psi}^*
\end{pmatrix}^\top$ is the Nambu spinor.
By construction, only the particle-hole symmetric Nambu spinors satisfying $C' \ket{\psi} = \ket{\psi}$ correspond to the spin fluctuations defined through Eq.~(\ref{eq:spin_wave_param}).
Using the eigenmodes $\ket{\phi_n^\mathrm{R}(0)}$ associated with the eigenvalues $E_n(0)$ of $\tilde{H}(0)$, $\ket{\psi}$ can be decomposed into Bogoliubov modes given by $\ket{\psi_n}=\ket{\phi_n^\mathrm{R}(0)} + C' \ket{\phi_n^\mathrm{R}(0)}$.
The excitation energy associated with the Bogoliubov mode $\ket{\psi_n}$ reads $\Delta E = E_n(0) \braket{\phi_n^\mathrm{R}(0)|\tau_z|\phi_n^\mathrm{R}(0)}$, which implies that the excitation of a particle mode with eigenenergy $E_n(0)$ is equivalent to that of a hole mode with eigenenergy $-E_n^*(0)$.
Note that $\braket{\phi_n^\mathrm{R}(0)|\tau_z|\phi_n^\mathrm{R}(0)} = 0$ for complex $E_n(0)$, indicating that dynamically unstable modes can grow without changing the energy.
Under an appropriate normalization, $\Delta E=E_n(0)$ for particle modes with real eigenenergy $E_n(0)$; hence, we regard the particle modes as physical modes.

Based on the above considerations, we now discuss thermodynamic stability conditions for the excitation energies around stationary states in the absence of damping.
To this end, we consider entirely real eigenvalues $E_n(0)$ of $\tilde{H}(0)$ and focus on the particle bands, satisfying $\braket{\phi_n^\mathrm{R}(0)|\tau_z|\phi_n^\mathrm{R}(0)}>0$.
If there exists a negative energy $E_n(0)<0$, the total energy of the system can decrease by exciting such negative-energy modes, which implies that the system is thermodynamically unstable.
Although this excitation is forbidden in isolated systems due to energy conservation, it can occur in open systems through intrinsic dissipation processes, e.g., Gilbert damping in spin systems.
Thus, for stable stationary spin states, all the excitation energies for particle bands must be nonnegative.
This conclusion can also be seen explicitly from the perturbative calculation of the eigenenergies in the presence of damping in the next subsection.

\subsection{Perturbative analysis of magnon eigenvalues}
We perturbatively calculate the magnon eigenvalues of Eq.~(\ref{eq:magnon_nh_hamiltonian}) in the presence of damping by assuming $|\alpha_{\bm{r},\bm{r}'}| \ll 1$.
Within this assumption, the non-Hermitian BdG Hamiltonian matrix (\ref{eq:magnon_nh_hamiltonian}) can be expanded by convergent power series for $\|\alpha\|<1$ as
\begin{align}
    \tilde{H}(\alpha) &= \tilde{H}(0) + \sum_{l=1}^{\infty} (-\ii \tau_z A)^l \tilde{H}(0).
\end{align}
We consider the eigenvalue problem
\begin{align}
    \tilde{H}(\alpha) \ket{\phi_n^\mathrm{R}(\alpha)} &= E_n(\alpha) \ket{\phi_n^\mathrm{R}(\alpha)}, \\
    \tilde{H}^\dagger(\alpha) \ket{\phi_n^\mathrm{L}(\alpha)} &= E_n^*(\alpha) \ket{\phi_n^\mathrm{L}(\alpha)},
\end{align}
where $\ket{\phi_n^\mathrm{R}(\alpha)}$ and $\ket{\phi_n^\mathrm{L}(\alpha)}$ are the right and left eigenstates of $\tilde{H}(\alpha)$, normalized as $\braket{\phi_n^\mathrm{L}(\alpha)|\phi_n^\mathrm{R}(\alpha)}=1$.
We consider a nondegenerate eigenvalue $E_n(0)$ of the unperturbed Hamiltonian $\tilde{H}(0)$ and calculate its continuation $E_n(\alpha)$ for $\tilde{H}(\alpha)$.
In the following, we treat $E_n(0)$ as complex in general, and later specialize to the real case.

According to nondegenerate perturbation theory, $E_n(\alpha)$, $\ket{\phi_n^\mathrm{R}(\alpha)}$, and $\ket{\phi_n^\mathrm{L}(\alpha)}$ can be expanded in powers of $\alpha$ for sufficiently small $\|\alpha\|$~\cite{katoPerturbationTheoryLinear1995}.
For simplicity, we use this fact without proof;
the proof and more general discussions without assuming nondegeneracy are given in Ref.~\cite{katoPerturbationTheoryLinear1995}.
As a result, we obtain the first-order correction to the eigenvalue $E_n(0)$ as
\begin{align}
    E_n(\alpha) &= E_n(0) \left(1 - \ii \alpha_{\mathrm{eff},n}\right) + O(\|\alpha\|^2), \label{eq:perturbed_energy}
\end{align}
where $\alpha_{\mathrm{eff},n}$ is an effective damping for the eigenmode $\ket{\phi_n^\mathrm{R}(\alpha)}$ defined by
\begin{align}
    \alpha_{\mathrm{eff},n} &= \braket{\phi_n^\mathrm{L}(0)|\tau_z A|\phi_n^\mathrm{R}(0)}. \label{eq:eff_damping}
\end{align}

Here, we assume that $E_n(0)$ is nondegenerate, and hence so is its complex conjugate $E_n^*(0)$.
Defining $n^*$ as the index associated with the eigenvalue $E_n^*(0)$, i.e., $E_{n^*}(0)=E_n^*(0)$, we find (see Appendix~\ref{sec:left_derivation} for details)
\begin{align}
    \tau_z \ket{\phi_{n^*}^\mathrm{R}(0)} &= \ket{\phi_{n}^\mathrm{L}(0)} \braket{\phi_{n}^\mathrm{R}(0)|\tau_z|\phi_{n^*}^\mathrm{R}(0)}, \label{eq:left_right_relation}
\end{align}
which can be used to remove $\ket{\phi_n^\mathrm{L}(0)}$ from Eq.~(\ref{eq:eff_damping}) as
\begin{align}
    \alpha_{\mathrm{eff},n} &= \frac{\braket{\phi_{n^*}^\mathrm{R}(0)|A|\phi_n^\mathrm{R}(0)}}{\braket{\phi_{n^*}^\mathrm{R}(0)|\tau_z|\phi_n^\mathrm{R}(0)}}.
\end{align}

Here, we rederive the thermodynamic stability condition in the presence of damping, mentioned at the previous subsection.
For $n=n^*$ corresponding to real eigenenergy $E_n(0)$, i.e., without dynamical instability, $\alpha_{\mathrm{eff},n}$ becomes real.
For a stationary spin state to be stable, $\Imag E_n(\alpha) = - E_n(0) \alpha_{\mathrm{eff},n}$ must be nonpositive for all $n$.
Since $A\succeq 0$, it follows that $E_n(0)$ and $\braket{\phi_n(0)|\tau_z|\phi_n(0)}$ have the same sign.
This coincides with the thermodynamic stability condition that particle (hole) modes, satisfying $\braket{\phi_n(0)|\tau_z|\phi_n(0)} > 0$ ($\braket{\phi_n(0)|\tau_z|\phi_n(0)} < 0$), have nonnegative (nonpositive) eigenvalues.

We now consider a periodic spin structure and decompose the lattice position as $\bm{r}=\bm{R}+\bm{r}_i$, where $\bm{R}$ denotes a Bravais lattice vector for magnetic unit cells and $\bm{r}_i$ denotes the position of the crystalline unit cell within the magnetic unit cell.
Performing the Fourier transform with respect to $\bm{R}$ under the assumption of translation symmetry, we obtain
\begin{align}
    E_{n\bm{k}}(\alpha) &= E_{n\bm{k}}(0) \left(1 - \ii \alpha_{\mathrm{eff},n\bm{k}}\right) + O(\|\alpha\|^2), \label{eq:perturbed_energy_k} \\
    \alpha_{\mathrm{eff},n\bm{k}} &= \frac{\braket{\phi_{n^* \bm{k}}^\mathrm{R}(0)|A_{\bm{k}}|\phi_{n\bm{k}}^\mathrm{R}(0)}}{\braket{\phi_{n^* \bm{k}}^\mathrm{R}(0)|\tau_z|\phi_{n\bm{k}}^\mathrm{R}(0)}}, \label{eq:eff_damping_k}
\end{align}
where $\bm{k}$ is the Bloch momentum and $n$ denotes the band index.
Here, the matrix element of $A_{\bm{k}}$ is given by $A_{\bm{k}, ij} = \sum_{\bm{R}} A_{\bm{R} + \bm{r}_i, \bm{r}_j} e^{-\ii \bm{k}\cdot(\bm{R} + \bm{r}_i - \bm{r}_j)}$.

In what follows, we further assume that (i) the nonlocal damping $\alpha_{\bm{R}+\bm{r}_i,\bm{R}'+\bm{r}_j}$ only couples lattice sites with the same $\bm{r}_i=\bm{r}_j$, and (ii) the coupling coefficient is independent of the sublattice position $\bm{r}_i$, i.e., $\alpha_{\bm{R}+\bm{r}_i,\bm{R}'+\bm{r}_j}=\alpha_{\bm{R}-\bm{R}',\bm{0}}\delta_{ij}$.
Assumption (i) makes $A_{\bm{k}}$ diagonal, and assumption (ii) makes it proportional to the identity.
These assumptions, like the scalar treatment of $\alpha_{\bm{r},\bm{r}'}$, are simplifications for making the effective damping analytically tractable.
Both assumptions hold for the nonlocal damping considered in Sec.~\ref{sec:result_magnon}, which couples neighboring lattice sites only along the $z$ direction, where the skyrmion-string lattice is uniform.
In this case, $A_{\bm{k}}(\succeq 0)$ is proportional to the identity and can thus be written as a nonnegative scalar $\alpha_{\bm{k}} \coloneqq \sum_{\bm{R}} \alpha_{\bm{R}, \bm{0}} e^{-\ii \bm{k}\cdot\bm{R}}\geq 0$, leading to
\begin{align}
    \alpha_{\mathrm{eff},n\bm{k}} &= \alpha_{\bm{k}} \eta_{n\bm{k}}, \label{eq:eff_damping_k_scalar} \\
    \eta_{n\bm{k}} &= \frac{\braket{\phi_{n^* \bm{k}}^\mathrm{R}(0)|\phi_{n\bm{k}}^\mathrm{R}(0)}}{\braket{\phi_{n^* \bm{k}}^\mathrm{R}(0)|\tau_z|\phi_{n\bm{k}}^\mathrm{R}(0)}}. \label{eq:tau_exp}
\end{align}

From now on, we assume that all eigenenergies $E_{n\bm{k}}(0)$ are real.
In this case, $\eta_{n\bm{k}} = \braket{\phi_{n \bm{k}}^\mathrm{R}(0)|\phi_{n\bm{k}}^\mathrm{R}(0)} / \braket{\phi_{n \bm{k}}^\mathrm{R}(0)|\tau_z|\phi_{n\bm{k}}^\mathrm{R}(0)}$ becomes real and represents the relative weights of the particle and hole components of the magnons.
In terms of spin dynamics, $\eta_{n\bm{k}}$ can be interpreted as a parameter characterizing the ellipticity of the spin trajectory~\cite{rozsaEffectiveDampingEnhancement2018,okumuraInstabilityMagneticSkyrmion2023,wangConnectionSpinwavePolarization2025}.
By the Cauchy-Schwarz inequality, $|\eta_{n\bm{k}}|\geq 1$, and $\eta_{n\bm{k}}=\pm 1$ is satisfied if and only if $\ket{\phi_{n\bm{k}}^\mathrm{R}(0)}$ is an eigenstate of $\tau_z$, i.e., $\tau_z \ket{\phi_{n\bm{k}}^\mathrm{R}(0)} = \pm \ket{\phi_{n\bm{k}}^\mathrm{R}(0)}$, which is the case when the particle space and the hole space are completely decoupled.
It follows that depending on the value of $\eta_{n\bm{k}}$ the damping may be effectively enhanced but never reduced below $\alpha_{\bm{k}}$.

Equations~(\ref{eq:perturbed_energy_k}) and (\ref{eq:eff_damping_k_scalar}) tell us that the spectrum can form a loop in the complex-energy plane when $\alpha_{\bm{k}}$ or $\eta_{n\bm{k}}$ has a nontrivial $\bm{k}$ dependence.
Although the presence of nonlocal damping is essential to satisfy the former condition, it is not essential for the latter one.

\section{\label{sec:winding_formula}Relation between spectral winding numbers and band minima}

\subsection{Winding number for point gaps}
Here, we briefly review the relation between the point-gap topology and the NHSE.
A Hamiltonian $\tilde{H}$ is called point-gapped at $E\in\mathbb{C}$ if the spectrum of $\tilde{H}$ does not contain $E$, i.e., $\det(\tilde{H} - E)\neq 0$~\cite{gongTopologicalPhasesNonHermitian2018,kawabataSymmetryTopologyNonHermitian2019}.
In the following, we single out a specific direction, say $x_\parallel$, and consider the Bloch Hamiltonian $\tilde{H}_{k_\parallel}$ parametrized by the Bloch momentum $k_\parallel$ under the periodic boundary condition (PBC) along the $x_\parallel$ direction.
When $\tilde{H}_{k_\parallel}$ has a point gap at $E$, one can define a spectral winding number as~\cite{gongTopologicalPhasesNonHermitian2018,kawabataSymmetryTopologyNonHermitian2019,okumaTopologicalOriginNonHermitian2020,zhangCorrespondenceWindingNumbers2020}
\begin{align}
    W(E) &= \frac{1}{2\pi\ii} \int_{-\pi/\Lambda}^{\pi/\Lambda} dk_\parallel\, \frac{\partial}{\partial k_\parallel} \log\det(\tilde{H}_{k_\parallel} - E), \label{eq:winding_number}
\end{align}
where $\Lambda$ is the period of the magnetic structure along the $x_\parallel$ direction.
It can be decomposed into the sum of the winding numbers for each band $E_{n k_\parallel}$ as $W(E) = \sum_{n} W_n(E)$ with
\begin{align}
    W_n(E) &= \frac{1}{2\pi\ii} \int_{-\pi/\Lambda}^{\pi/\Lambda} dk_\parallel\, \frac{\partial}{\partial k_\parallel} \log(E_{n k_\parallel} - E). \label{eq:spectral_winding_number}
\end{align}
According to Refs.~\cite{okumaTopologicalOriginNonHermitian2020,zhangCorrespondenceWindingNumbers2020}, the NHSE occurs under the open boundary condition along the $x_\parallel$ direction ($x_\parallel$-OBC) if and only if there exists $E$ such that $W(E)\neq 0$ under the PBC.
The skin mode with eigenenergy $E$ and $W(E)=-1$ ($W(E)=+1$) is localized at the boundary in the $+x_\parallel$ ($-x_\parallel$) direction, which we call the top (bottom) boundary.
We note that not all the eigenmodes with eigenenergy $E$ residing in the interior of the PBC spectrum necessarily become the skin modes under the $x_\parallel$-OBC with finite system size (see, e.g., Refs.~\cite{okumaTopologicalOriginNonHermitian2020,okumaNonHermitianTopologicalPhenomena2023} for further details on these subtleties).

\subsection{Evaluation of winding numbers for local and nonlocal damping}
We evaluate Eq.~(\ref{eq:spectral_winding_number}) for the BdG Hamiltonian (\ref{eq:magnon_nh_hamiltonian}) by using the analytical expression for the eigenenergy (\ref{eq:perturbed_energy_k}) obtained through the perturbation theory.
In the derivation, we assume that the band dispersion in the absence of damping, $E_{n k_\parallel}(0)$, is real and has a unique minimum $E_0$ at $k_\parallel=k_0$.
In addition, we assume that $E_{n k_\parallel}(0)$ does not cross the other bands around $k_\parallel=k_0$, so that Eq.~(\ref{eq:perturbed_energy_k}) can be used in this region.
Within these assumptions, we can show that there exists $E$ such that
\begin{align}
    W_n(E) &= \sgn E_0 \sgn\left.\partial_{k_\parallel}^{2l_0 + 1} \alpha_{\mathrm{eff},n k_\parallel}\right|_{k_\parallel=k_0}, \label{eq:winding_formula}
\end{align}
with $l_0$ being the smallest nonnegative integer such that $\left.\partial_{k_\parallel}^{2l_0 + 1} \alpha_{\mathrm{eff},n k_\parallel}\right|_{k_\parallel=k_0} \neq 0$.
Here, we assign $\sgn E_0=+1$ for $E_0=0$ (see Appendix~\ref{sec:winding_derivation} for details).

To further proceed, we need a concrete expression for $\alpha_{\mathrm{eff},n k_\parallel}=\alpha_{k_\parallel} \eta_{n k_\parallel}$.
Below, we assume $l_0=0$ and consider two special cases in which we can separate contributions from two terms in $\partial_{k_\parallel} \alpha_{\mathrm{eff},n k_\parallel} = \alpha_{k_\parallel} \partial_{k_\parallel} \eta_{n k_\parallel} + \eta_{n k_\parallel} \partial_{k_\parallel} \alpha_{k_\parallel}$:
(i) $\alpha_{k_\parallel}$ is almost unchanged around $k_\parallel=k_0$, i.e., $\partial_{k_\parallel} \alpha_{\mathrm{eff},n k_\parallel} \approx \alpha_{k_\parallel} \partial_{k_\parallel} \eta_{n k_\parallel}$ at $k_\parallel \approx k_0$, and (ii) $\eta_{n k_\parallel}$ is almost unchanged around $k_\parallel=k_0$, i.e., $\partial_{k_\parallel} \alpha_{\mathrm{eff},n k_\parallel} \approx \eta_{n k_\parallel} \partial_{k_\parallel} \alpha_{k_\parallel}$ at $k_\parallel \approx k_0$.

Case (i) is exactly satisfied when nonlocal damping is absent, as in the conventional LLG equation with $\alpha_{\bm{r},\bm{r}'}=\alpha_0 \delta_{\bm{r},\bm{r}'}$.
In case (i), nontrivial spectral windings are attributed to the nontrivial $k_\parallel$ dependence of $\eta_{n k_\parallel}$ since $\alpha_{k_\parallel} \geq 0$.
Although $\sgn \partial_{k_\parallel} \eta_{n k_\parallel}$ cannot be calculated unless a concrete expression for $\eta_{n k_\parallel}$ is known, if $\eta_{n k_\parallel}$ is monotonic around $k_\parallel=k_0$, then $W_n(E)\neq 0$ is satisfied for some $E$.

Case (ii) is exactly satisfied for field-polarized states, where particle and hole spaces are completely decoupled.
In case (ii), we have to consider only $\sgn \partial_{k_\parallel} \alpha_{k_\parallel}$ because $\sgn \eta_{n k_\parallel}$ can be fixed by choosing either particle or hole bands.
In the following, we employ the concrete model of the damping given by
\begin{align}
    \alpha_{k_\parallel} &= \alpha_0 + 2\alpha_\parallel \cos(k_\parallel \Lambda), \label{eq:damping}
\end{align}
which represents local ($\alpha_0 > 0$) and nonlocal ($\alpha_\parallel > 0$) damping terms along the $x_\parallel$ direction.
The nonlocal damping couples the sites displaced by $\pm\Lambda$ from each other, i.e., the corresponding sites in the nearest-neighbor magnetic unit cells.
In this case, since $\sgn \partial_{k_\parallel} \alpha_{k_\parallel}=-\sgn k_\parallel$ for $k_\parallel \in (-\pi/\Lambda, 0) \cup (0, \pi/\Lambda)$ and $\sgn \partial_{k_\parallel} \alpha_{k_\parallel}=0$ otherwise, there exists $E$ such that $W_n(E)\neq 0$, where
\begin{align}
	W_n(E) &= -\sgn k_0 \sgn E_0 \sgn \eta_{n k_\parallel}, \label{eq:winding_local_minimum}
\end{align}
only if $k_0\neq 0,\pm\pi/\Lambda$.
Here, $\sgn E_0 = 1$ and $\sgn \eta_{n k_\parallel} = 1$ for particle bands in thermodynamically stable stationary states (see Sec.~\ref{subsec:particle_hole}).

\section{\label{sec:result_magnon}Numerical evaluation of nontrivial winding numbers in skyrmion strings}

\subsection{\label{sec:model}Model}
To demonstrate the results obtained in Sec.~\ref{sec:winding_formula}, we consider an electron spin model ($\gamma<0$) on a triangular lattice in $xy$ plane spanned by primitive translation vectors $\bm{a}_1=(a, 0, 0)$, $\bm{a}_2=(-a/2, \sqrt{3}a/2, 0)$, $\bm{a}_3=(0, 0, c)$.
We will discuss the NHSE along the $x_\parallel=z$ direction.
The Hamiltonian is given by
\begin{align}
    \mathcal{H} &= \frac{1}{2} \sum_{\bm{r}, \bm{r}'} \left[-J_{\bm{r}, \bm{r}'} \bm{S}_{\bm{r}} \cdot \bm{S}_{\bm{r}'} + \bm{D}_{\bm{r}, \bm{r}'} \cdot (\bm{S}_{\bm{r}} \times \bm{S}_{\bm{r}'})\right] \notag \\
    &\quad - \sum_{\bm{r}} \hbar\gamma \bm{B}_{\bm{r}} \cdot \bm{S}_{\bm{r}}, \label{eq:spin_hamiltonian}
\end{align}
where we consider only nearest-neighbor interactions: $J_{\bm{r}, \bm{r}'} = J_{\bm{r}', \bm{r}}$ is a nearest-neighbor exchange interaction (or Ruderman-Kittel-Kasuya-Yosida (RKKY) interaction) and $\bm{D}_{\bm{r}, \bm{r}'} = -\bm{D}_{\bm{r}', \bm{r}}$ is a nearest-neighbor Dzyaloshinskii-Moriya (DM) vector.
$\bm{B}_{\bm{r}}$ is an external magnetic field.
For the spin Hamiltonian~(\ref{eq:spin_hamiltonian}), the blocks of the Hessian matrix~(\ref{eq:BdG_hamiltonian}) read
\begin{align}
    h_{\bm{r}, \bm{r}'} &= S \left[-J_{\bm{r}, \bm{r}'} \tilde{\bm{e}}_{\bm{r}}^+ \cdot \tilde{\bm{e}}_{\bm{r}'}^- + \bm{D}_{\bm{r}, \bm{r}'} \cdot (\tilde{\bm{e}}_{\bm{r}}^+ \times \tilde{\bm{e}}_{\bm{r}'}^-)\right] - \delta_{\bm{r}, \bm{r}'} \lambda_{\bm{r}}, \\
    \lambda_{\bm{r}} &= S \sum_{\bm{r}''} \left[-J_{\bm{r}, \bm{r}''} \tilde{\bm{e}}_{\bm{r}}^z \cdot \tilde{\bm{e}}_{\bm{r}''}^z + \bm{D}_{\bm{r}, \bm{r}''} \cdot (\tilde{\bm{e}}_{\bm{r}}^z \times \tilde{\bm{e}}_{\bm{r}''}^z)\right] \notag \\
    &\quad - \hbar\gamma \tilde{\bm{e}}_{\bm{r}}^z \cdot \bm{B}_{\bm{r}}, \\
    \Delta_{\bm{r}, \bm{r}'} &= S \left[-J_{\bm{r}, \bm{r}'} \tilde{\bm{e}}_{\bm{r}}^+ \cdot \tilde{\bm{e}}_{\bm{r}'}^+ + \bm{D}_{\bm{r}, \bm{r}'} \cdot (\tilde{\bm{e}}_{\bm{r}}^+ \times \tilde{\bm{e}}_{\bm{r}'}^+)\right].
\end{align}

We assume the Bloch type DM interaction, $D_{\bm{r}, \bm{r}'}\propto \bm{r} - \bm{r}'$.
We also assume anisotropic coupling strengths;
the magnitude for $J_{\bm{r},\bm{r}'}, \bm{D}_{\bm{r},\bm{r}'}$ is given by $J, D$ when $\bm{r} - \bm{r}'$ lies in the $xy$ plane and $J_\parallel, D_\parallel$ when $\bm{r} - \bm{r}'$ is along the $z$ direction, respectively.
We consider a uniform external magnetic field along the $z$ direction, $\bm{B}_{\bm{r}}=B \bm{e}_z$, and define a unit of magnetic field as $b\coloneqq JS/\hbar|\gamma|$.

We use a momentum-dependent damping coefficient given by Eq.~(\ref{eq:damping}).
In the following sections, we consider two cases: (i) purely local damping and (ii) additional nonlocal damping along the $z$ direction.
The former case models three-dimensional systems in the bulk, and the latter case models stacked multilayer systems along the $z$ direction.

For field-polarized states with $\tilde{\bm{e}}_{\bm{r}}^z=\sgn\gamma \bm{e}_z$, there is only a single magnon particle band with dispersion $E_{k_z}(0) = 2S \{J_\parallel [1 - \cos(k_z c)] + D_\parallel \sin(k_z c)\} + \hbar |\gamma| B$~\cite{liNonHermitianLiouvillianSkin2025}, which determines the localized position of the skin mode, fixed by $\sgn k_0=-\sgn(D_\parallel/J_\parallel)$ via Eq.~(\ref{eq:winding_local_minimum}).
This is not the case for, e.g., skyrmion-string lattices, whose spin direction varies in space, and indeed, the existence of the bands hosting the skin modes localized at top and bottom boundaries can be seen in Fig.~\ref{fig:eigenvalue_kz_low-B}.

\begin{figure}
    \includegraphics[width=0.8\linewidth]{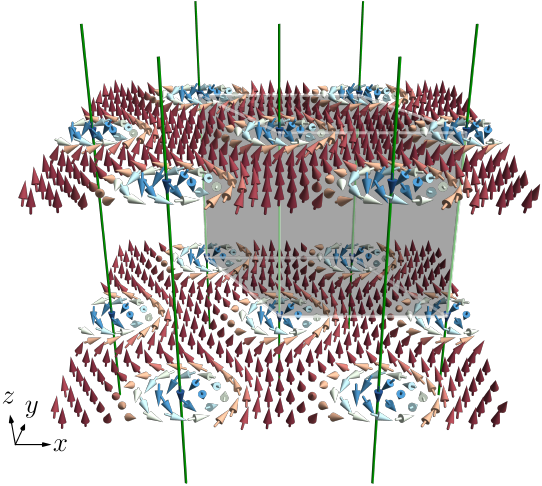}
    \caption{\label{fig:spin_config}Spin configuration of the stationary skyrmion-string-lattice state at $B/b=0.7$ with $b\coloneqq JS/\hbar|\gamma|$.
    We fix $D/J=D_\parallel/J_\parallel=1$ and $J_\parallel/J=1/2$.
    Each arrow at site $\bm{r}$ represents a direction of the magnetic moment $\gamma \bm{S}_{\bm{r}}$ of electrons ($\gamma<0$).
    Green lines indicate the center of the skyrmions.
    The magnetic unit cell is shown as a gray parallelepiped.}
\end{figure}

In the following subsections, we examine the NHSE in stationary skyrmion-string lattice states in detail.
Such stationary states are obtained by relaxing the superposition of triple helices in the $xy$ plane, uniform along the $z$ direction, via the LLG equation.
In other words, the magnetic period along the $z$ direction for the uniform skyrmion strings coincides with the lattice constant, $\Lambda = c$.
We consider systems composed of a single magnetic unit cell containing $9\times 9\times 1$ spins with PBCs imposed in the in-plane directions.
Accordingly, our analysis is restricted to the magnon modes with zero in-plane momentum, $k_x=k_y=0$, and the spectral winding is considered with respect to $k_z$.
As we explain in Sec.~\ref{sec:propagation}, this restriction is sufficient to describe the propagation dynamics considered in this work.
We fix $D/J=D_\parallel/J_\parallel=1$ and $J_\parallel/J=1/2$.
The stationary spin state at $B/b=0.7$ is shown in Fig.~\ref{fig:spin_config}.
We note that the obtained stationary states of the skyrmion strings are not necessarily ground states.
However, such metastable states can be realized by quenching the external magnetic field and temperature~\cite{kagawaCurrentinducedViscoelasticTopological2017}.

\subsection{\label{sec:magnon_spectrum}Magnon spectrum}
\begin{figure}
    \includegraphics[width=\linewidth]{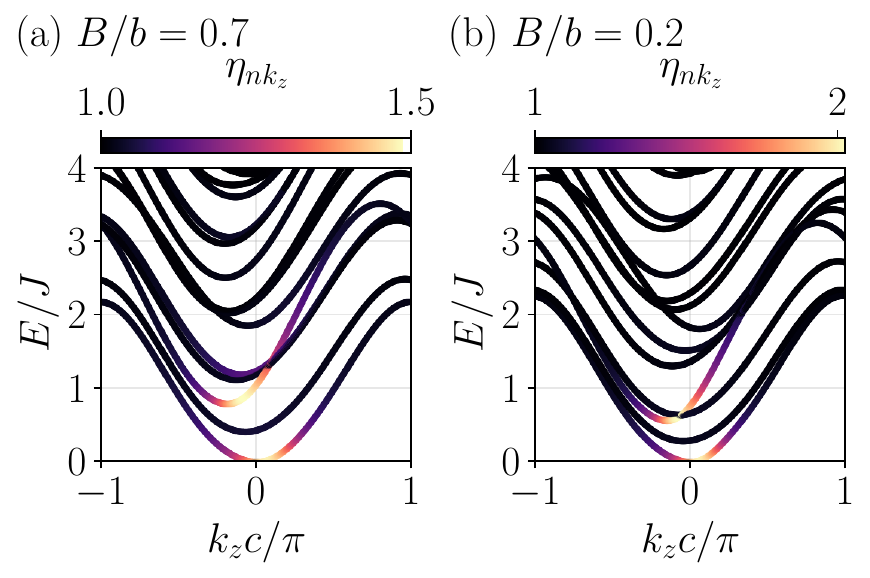}
    \caption{\label{fig:eigenvalue_kz_real}Magnon band structures without damping along the momentum $\bm{k} = (0, 0, k_z)$ in the Brillouin zone.
    The magnetic field is (a) $B/b=0.7$ and (b) $B/b=0.2$.
    The color indicates $\eta_{n k_z}$ defined by Eq.~(\ref{eq:tau_exp}) for each eigenstate.}
\end{figure}

We calculate the magnon band structures in the absence of damping by diagonalizing $\tilde{H}(0)=\tau_z H$.
Figure~\ref{fig:eigenvalue_kz_real} shows the obtained low-energy magnon band structures at (a) $B/b=0.7$ and (b) $B/b=0.2$ along the momentum $\bm{k} = (0, 0, k_z)$ in the Brillouin zone.
Since the excitation spectra are real and nonnegative, the underlying stationary spin states are stable.
At a high magnetic field with $B/b=0.7$, all the bands reach their minima at $k_z\leq 0$.
In contrast, at a low magnetic field with $B/b=0.2$, some bands reach their minima at positive momenta; in particular the seventh lowest-energy band does so [see Fig.~\ref{fig:eigenvalue_kz_real}(b)].
We have confirmed that, in the presence of the small damping (e.g., $\alpha_0=0.1$ and $\alpha_\parallel=0.05$), the real parts of the eigenenergies remain almost unchanged on the scale shown in Fig.~\ref{fig:eigenvalue_kz_real}, thereby supporting the validity of the perturbatively obtained eigenenergy in Eq.~(\ref{eq:perturbed_energy_k}).

The color in Fig.~\ref{fig:eigenvalue_kz_real} indicates $\eta_{n k_z}$ (\ref{eq:tau_exp}) evaluated for each eigenstate at $k_z$.
Two modes exhibit a strong $k_z$ dependence: the lowest-energy band ($n=1$) and the third lowest-energy band ($n=3$).
The former mode corresponds to a Nambu-Goldstone mode associated with the spontaneously broken translation symmetry in the skyrmion-string lattice~\cite{petrovaSpinWavesSkyrmion2011}, in which the center of mass of the skyrmions oscillates clockwise around the $z$ axis.
The latter mode, which is the main focus of our interest in the following calculations, is dubbed a counterclockwise (CCW) mode~\cite{mochizukiSpinWaveModesTheir2012,sekiPropagationDynamicsSpin2020}, characterized by CCW motion of the skyrmion center of mass.
Similarly, the sixth (seventh) lowest-energy band at $B/b=0.7$ ($B/b=0.2$) is dubbed a clockwise (CW) mode~\cite{mochizukiSpinWaveModesTheir2012,sekiPropagationDynamicsSpin2020}, characterized by CW motion of the skyrmion center of mass.
The $\eta_{n k_z}$ for the CCW mode deviates from 1 as a function of $k_z$ and reaches its maximum at a momentum larger than $k_0$, the point where the energy band reaches its minimum.
It follows that $\eta_{n k_z}$ monotonically increases in the neighborhood of $k_z=k_0$.
Therefore, in the presence of local damping, this mode is expected to exhibit a nontrivial winding $W_3(E)=+1$ under the PBC and to show the NHSE under the $z$-OBC.
We demonstrate this explicitly below.
We note that, at $\alpha=0$, no boundary-localized modes appear under the $z$-OBC even though the chiral skyrmion texture breaks the symmetry between the $\pm z$ directions: The spectrum is entirely real, and the system is topologically trivial with respect to both point and line gaps.
We have also confirmed that all eigenmodes remain extended.
The localization discussed below is therefore entirely induced by the damping.

\begin{figure}
    \includegraphics[width=\linewidth]{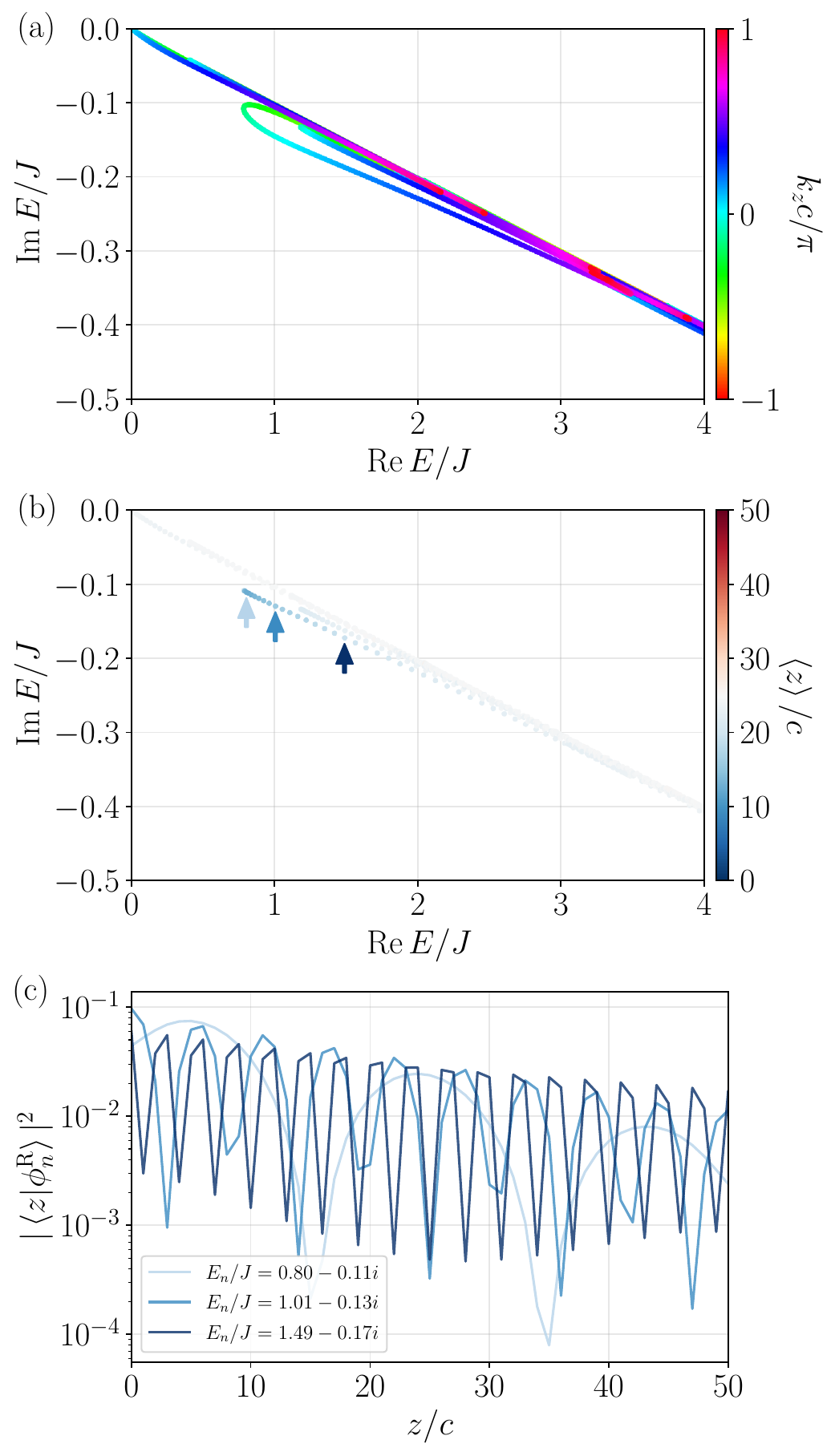}
    \caption{\label{fig:eigenvalue_kz}(a), (b) Magnon complex spectra at $B/b=0.7$ with local damping only ($\alpha_0=0.1, \alpha_\parallel=0$),
    (a) PBC; (b) OBC along the $z$ direction with $N_z = 51$ sites along $z$.
    In (a), colors denote the momentum $k_z$.
    In (b), colors denote $\braket{z} \coloneqq \braket{\phi_n^\mathrm{R}|z|\phi_n^\mathrm{R}} / \braket{\phi_n^\mathrm{R}|\phi_n^\mathrm{R}}$, where $\ket{\phi_n^\mathrm{R}}$ is the right eigenstate.
    (c) Spatial distribution of the squared amplitude of representative skin modes indicated by the arrows in (b).}
\end{figure}

We now discuss the effect of damping.
We first consider case (i), where only local damping is present, i.e., $\alpha_0\neq 0$ and $\alpha_\parallel=0$.
Figures~\ref{fig:eigenvalue_kz}(a) and \ref{fig:eigenvalue_kz}(b) show the complex magnon spectra at $B/b=0.7$ and $\alpha_0=0.1$ under (a) the PBC and (b) the $z$-OBC.
The spectrum acquires imaginary parts due to the nonzero local damping, but almost all bands do not show loop structures in the complex-energy plane.
Among them, only the spectrum for the CCW mode forms a loop with winding $W_3(E)=+1$, which is satisfied for $E$ at the interior of the loop [see Fig.~\ref{fig:eigenvalue_kz}(a)].
As expected from the behavior of the complex PBC spectrum, most bands remain extended Bloch waves, whereas the CCW mode and some other modes with windings become the skin modes localized at the bottom boundary under the $z$-OBC (see Fig.~\ref{fig:eigenvalue_kz}(b), where the color indicates the expectation value of the position of the eigenmodes along the $z$ direction).
Figure~\ref{fig:eigenvalue_kz}(c) shows the spatial profiles of three skin modes originating from the CCW mode.
Apart from rapid oscillations, the profiles decay exponentially away from the bottom boundary.
Although the spectrum in the presence of $\alpha_0$ alone is shown only for $B/b=0.7$ here, the overall spectral structure is almost the same for $B/b=0.2$.

\begin{figure}
    \includegraphics[width=\linewidth]{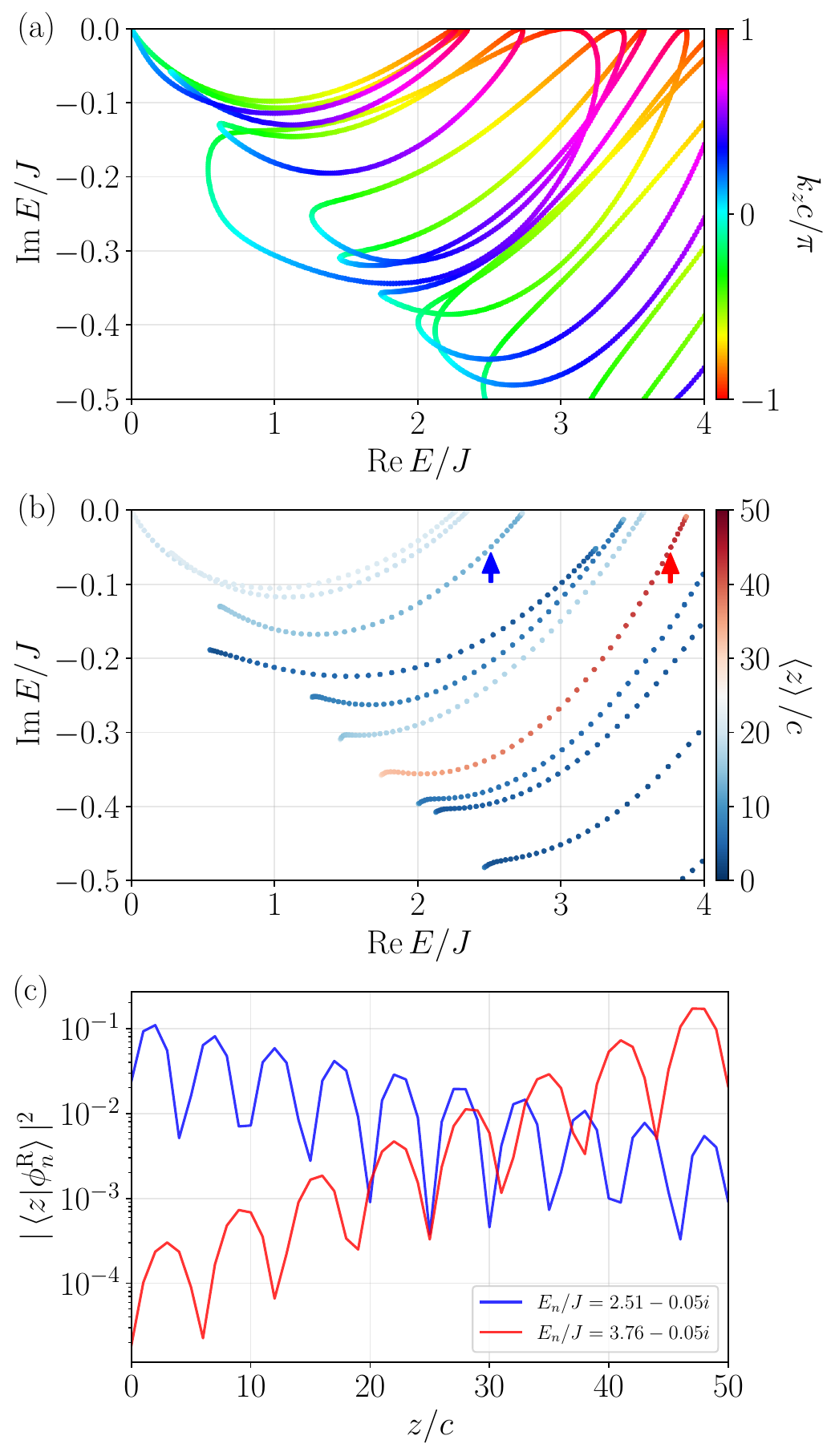}
    \caption{\label{fig:eigenvalue_kz_low-B}(a), (b) Magnon complex spectra at $B/b=0.2$ with nonlocal damping ($\alpha_0 = 0.1, \alpha_\parallel = 0.05$).
    (a) and (b) correspond to the PBC and the OBC along $z$ ($N_z = 51$ sites), respectively.
    Color coding follows that in Fig.~\ref{fig:eigenvalue_kz}.
    (c) Spatial distribution of the squared amplitude of representative skin modes indicated by the arrows in (b).}
\end{figure}

Next, we consider case (ii), where both local and nonlocal damping coexist, based on Eq.~(\ref{eq:winding_local_minimum}).
Since all the bands at $B/b=0.7$ reach their minimum at a negative momentum [see Fig.~\ref{fig:eigenvalue_kz_real}(a)], they have positive windings ($W_n(E)=+1$) and thus are localized at the bottom boundary, as is the case for ferromagnetic multilayers~\cite{liNonHermitianLiouvillianSkin2025}.
However, this is not the case for $B/b=0.2$ because the seventh lowest-energy band reaches its minimum at a positive momentum [see Fig.~\ref{fig:eigenvalue_kz_real}(b)], which gives a negative winding with $W_7(E)=-1$.
Figures~\ref{fig:eigenvalue_kz_low-B}(a) and \ref{fig:eigenvalue_kz_low-B}(b) show the magnon complex spectra at $B/b=0.2$ with $\alpha_0=0.1$ and $\alpha_\parallel=0.05$.
Due to the nonlocal damping, all bands exhibit nonzero windings, most of which are positive, resulting in the bottom-boundary-localized skin modes.
In contrast, the seventh lowest-energy band has a negative winding, leading to the top-boundary-localized skin modes.
The spatial profiles of these two types of skin modes are shown in Fig.~\ref{fig:eigenvalue_kz_low-B}(c).
The profile of a mode from a band with a positive winding ($E_n/J=2.51-0.05\ii$) decays exponentially away from the bottom boundary, whereas that of a mode from the seventh lowest-energy band ($E_n/J=3.76-0.05\ii$) decays away from the top boundary.
As shown above, the boundary for skin-mode localization can differ from band to band in multiband noncollinear spin systems, such as skyrmion-string lattices, even at fixed damping.
This band dependence is intrinsic to multiband systems.

Here, we comment on the applicability of the generalized Brillouin zone (GBZ) in the non-Bloch band theory~\cite{yaoEdgeStatesTopological2018,yokomizoNonBlochBandTheory2019}, which provides the localization length of each skin mode.
In the standard approach, the GBZ is obtained by solving the characteristic equation $\det(\tilde{H}_\beta - E) = 0$ as an algebraic equation for $\beta$, the analytic continuation of $e^{\ii k_z c}$, under the condition that the two middle roots ordered by their moduli share the same modulus~\cite{yokomizoNonBlochBandTheory2019}.
For our magnon Hamiltonian, however, this procedure is difficult for two reasons.
First, the magnetic unit cell contains $81$ spins, so that the BdG Hamiltonian has $162$ internal degrees of freedom and the degree of the characteristic polynomial reaches $2\times 162=324$ even for the nearest-neighbor couplings along the $z$ direction.
This large degree makes the numerical determination of the GBZ unstable.
Second, the exact symbolic computation of the GBZ requires the model parameters to be given exactly, e.g., as rational numbers~\cite{yokomizoNonBlochBandTheory2019,yangNonHermitianBulkBoundaryCorrespondence2020}, whereas the coefficients of our Hamiltonian are determined by the numerically obtained stationary spin configuration and are not available in a closed form.
We therefore rely on the point-gap winding number, which diagnoses the occurrence of the NHSE without computing the GBZ~\cite{okumaTopologicalOriginNonHermitian2020,zhangCorrespondenceWindingNumbers2020}.

\subsection{\label{sec:propagation}Spin-wave propagation in the presence of damping}
A nontrivial spectral winding number implies asymmetric damping between positive- and negative-group-velocity modes~\cite{besshoNielsenNinomiyaTheoremBulk2021}.
To detect such nontrivial topology in our system, we investigate spin-wave propagation in bulk by solving Eq.~(\ref{eq:generalized_LLG}) starting from the stationary skyrmion-string-lattice states.
We set the system size along the $z$ direction to $N_z=101$ and impose the PBC.
We apply an additional in-plane oscillating magnetic-field pulse at the central plane along the $z$ axis, $z=\lfloor N_z/2\rfloor c$:
\begin{align}
    \bm{B}_{\mathrm{osc},\bm{r}}^\pm(t) &= B_\mathrm{osc} f_\mathrm{e}(t) [\cos(\Omega t) \bm{e}_x \pm \sin(\Omega t) \bm{e}_y] \delta_{z, \lfloor\frac{N_z}{2}\rfloor c}, \label{eq:driving_magnetic_field}
\end{align}
where $f_\mathrm{e}(t) = e^{-(t - t_0)^2 / 2\sigma^2}$ is an envelope function, and the $+$ and $-$ signs correspond to left- and right-circularly rotating waves, respectively.
Since the driving pulse~(\ref{eq:driving_magnetic_field}) is spatially uniform within the $xy$ plane, it excites only the magnon modes with $k_x=k_y=0$.
Such an excitation is the simplest one and has been realized experimentally~\cite{sekiPropagationDynamicsSpin2020}.
Moreover, the stationary texture, the damping terms, and the boundary conditions all preserve the in-plane magnetic translation symmetry, so that the excited spin waves remain in the $k_x=k_y=0$ sector during the time evolution.
Therefore, the point-gap topology relevant to the propagation dynamics is the winding number $W(E)$ evaluated at $k_x=k_y=0$.
We numerically solve Eq.~(\ref{eq:generalized_LLG}) by the implicit midpoint method~\cite{luInfluenceNonlocalDamping2023}.
Parameters are fixed as $B_\mathrm{osc}/b=0.01$ and $t_0 = 5 \hbar/J$.
The total spin-wave amplitude (or magnon occupation number) along the $z$ axis, $n_z(t)$, is evaluated as the summation of $|\psi_{\bm{r}}(t)|^2$, defined by Eq.~(\ref{eq:spin_wave_param}), over the site $\bm{r}=(x, y, z)$ residing in a magnetic unit cell at fixed $z$.
We also calculate the magnon number $n_{k_z}(t)$ as a function of $k_z$ in the same way as $n_z(t)$ with $\psi_{\bm{r}}(t)$ replaced by its Fourier transform along the $z$ axis, $\psi_{k_z}(t)$.
Furthermore, we compute the magnon number $n_{k_z}(\omega)$ in the frequency domain through the Fourier transform of $\psi_{k_z}(t)$ in the time direction.

\begin{figure*}
    \includegraphics[width=0.9\linewidth]{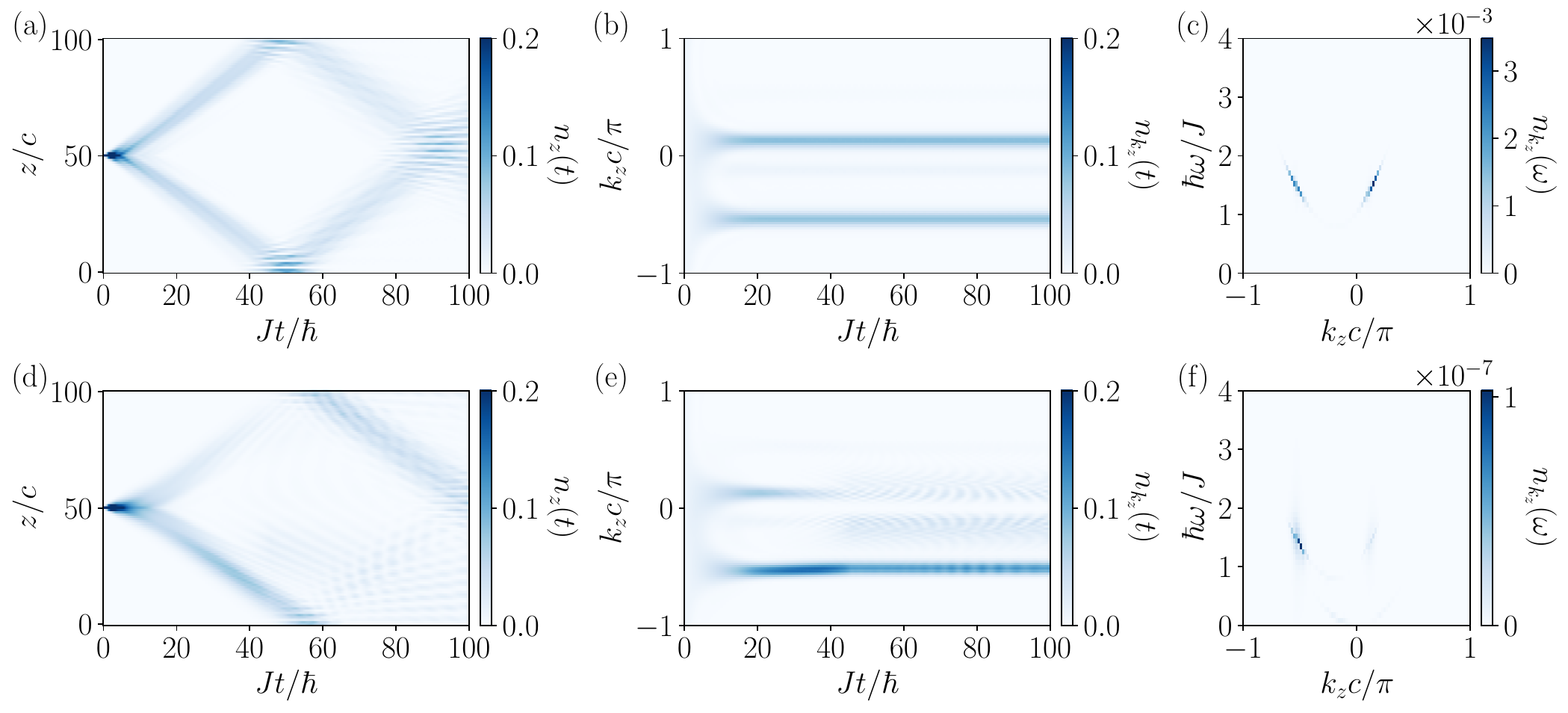}
    \caption{\label{fig:llg_magnon_z}Spin-wave propagation excited by a left-circularly rotating magnetic-field pulse $\bm{B}_{\mathrm{osc},\bm{r}}^+$ with frequency $\Omega=1.5J/\hbar$, applied only at the the single layer at $z = 50c$ under the PBC in all directions for a system with $N_z = 101$ sites along $z$ at $B/b=0.7$.
    The color indicates the magnon number, normalized at every time $t$ (a), (b), (d), (e).
    The pulse parameters are $B_\mathrm{osc}/b=0.01,t_0=5\hbar/J,\sigma=6\hbar/J$.
    The Gilbert damping is (a)--(c) $\alpha_0=\alpha_\parallel=0$ and (d)--(f) $\alpha_0=0.1$ and $\alpha_\parallel=0$.
    (a), (d) Time evolution of the magnon number along the $z$ axis.
    (b), (e) Corresponding time evolution in momentum space.
    (c), (f) Corresponding magnon number in momentum and frequency space, calculated from the data with $30 \leq Jt/\hbar \leq 100$.
    In (d)--(f), the downward-propagating mode ($k_z<0$) persists for a longer time, reflecting the nontrivial point-gap topological invariant $W_3(E)$.}
\end{figure*}

We apply the left-circularly rotating field $\bm{B}_{\mathrm{osc},\bm{r}}^+$ with frequency $\Omega=1.5J/\hbar$ to excite the CCW mode (the third lowest-energy mode) at $B/b=0.7$.
Figure~\ref{fig:llg_magnon_z} shows the spin-wave propagation excited by the driving pulse with $\sigma = 6\hbar/J$.
In the absence of damping, i.e., $\alpha_0=\alpha_\parallel=0$, both positive- and negative-group-velocity modes are excited, and the total magnon number remains conserved after the driving pulse ends, provided that its amplitude is sufficiently small [see Figs.~\ref{fig:llg_magnon_z}(a)--\ref{fig:llg_magnon_z}(c)].
In the presence of local damping ($\alpha_0=0.1$ and $\alpha_\parallel=0$), in contrast, upward-propagating magnons dissipate over time, whereas downward-propagating magnons persist longer [see Figs.~\ref{fig:llg_magnon_z}(d)--\ref{fig:llg_magnon_z}(f)].
The difference in their lifetime originates from the nontrivial point-gap topological invariant $W_3(E)$~\cite{besshoNielsenNinomiyaTheoremBulk2021}.

\begin{figure*}
    \includegraphics[width=0.9\linewidth]{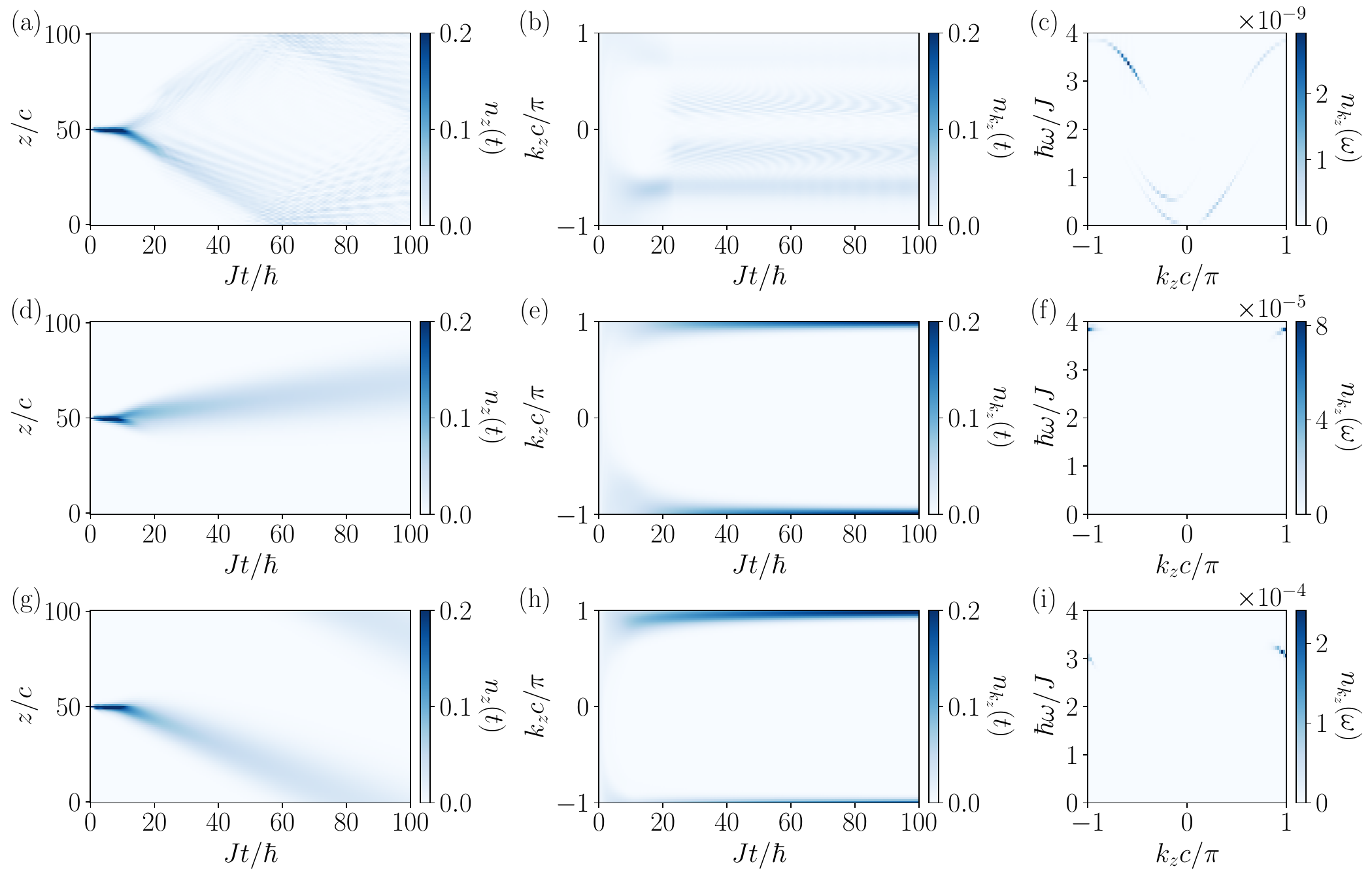}
    \caption{\label{fig:llg_magnon_z_low-B}Spin-wave propagation excited by a circularly rotating magnetic-field pulse with frequency $\Omega=3.5J/\hbar$, applied only at the single layer at $z=50c$ under the PBC at $B/b=0.2$.
    The rotating direction of the pulse is (a)--(f) right ($\bm{B}_{\mathrm{osc},\bm{r}}^-$) and (g)--(i) left ($\bm{B}_{\mathrm{osc},\bm{r}}^+$).
    The damping parameters are (a)--(c) $\alpha_0=0.1,\alpha_\parallel=0$ and (d)--(i) $\alpha_0=0.1,\alpha_\parallel=0.05$.
    (c), (f), and (i) are calculated from the data with $20 \leq Jt/\hbar \leq 100$.
    The other parameters are the same as Fig.~\ref{fig:llg_magnon_z}, except for the pulse parameter $\sigma=3\hbar/J$.}
\end{figure*}

To investigate the effect of nonlocal damping, we next focus on the CW mode (the seventh lowest-energy mode) at $B/b=0.2$.
Accordingly, we apply a right-circularly polarized magnetic field $\bm{B}_{\mathrm{osc},\bm{r}}^-$ with frequency $\Omega=3.5J/\hbar$.
Figures~\ref{fig:llg_magnon_z_low-B}(a)--\ref{fig:llg_magnon_z_low-B}(c) and \ref{fig:llg_magnon_z_low-B}(d)--\ref{fig:llg_magnon_z_low-B}(f) show the spin-wave propagation induced by this pulse in the absence ($\alpha_\parallel=0$) and presence ($\alpha_\parallel=0.05$) of nonlocal damping, respectively, with $\alpha_0=0.1$.
Initially, the downward-propagating mode is more strongly excited than the upward-propagating mode [see Figs.~\ref{fig:llg_magnon_z_low-B}(a) and \ref{fig:llg_magnon_z_low-B}(b)].
However, these modes rapidly decay, and off-resonant lower-energy modes become dominant for $Jt/\hbar\gtrsim 20$.
In contrast, in Figs.~\ref{fig:llg_magnon_z_low-B}(d)--\ref{fig:llg_magnon_z_low-B}(f), where nonlocal damping is present, only the upward-propagating mode survives.
This behavior is attributed to the nontrivial winding number $W_7(E)=-1$.
Interestingly, the propagation direction of magnons depends on the sign of the winding number and therefore differs depending on the excited mode.
Figures~\ref{fig:llg_magnon_z_low-B}(g)--\ref{fig:llg_magnon_z_low-B}(i) show the result obtained under the same conditions as in Figs.~\ref{fig:llg_magnon_z_low-B}(d)--\ref{fig:llg_magnon_z_low-B}(f), but the system is driven by a left-circularly polarized magnetic-field pulse $\bm{B}_{\mathrm{osc},\bm{r}}^+$.
In this case, the seventh CW mode is not excited; instead, the third CCW mode with $W_3(E)=+1$ is activated, and the magnons propagate exclusively in the downward direction.
In both Figs.~\ref{fig:llg_magnon_z_low-B}(e) and \ref{fig:llg_magnon_z_low-B}(h), the dynamics eventually become dominated by the mode at $k_z=\pi/c$ [see Figs.~\ref{fig:llg_magnon_z_low-B}(f) and \ref{fig:llg_magnon_z_low-B}(i)], which has the smallest $|\Imag E|$ [see Fig.~\ref{fig:eigenvalue_kz_low-B}(a)].

Finally, we discuss the relation among the chirality of the static skyrmion texture, the rotational character of the magnon modes, the winding numbers, and the direction of the asymmetric propagation.
We have numerically confirmed that reversing the sign of the DM interaction, $\bm{D}_{\bm{r},\bm{r}'}\to-\bm{D}_{\bm{r},\bm{r}'}$, i.e., $(D, D_\parallel)\to(-D, -D_\parallel)$, which reverses the chirality of the skyrmion-string lattice, maps the magnon dispersion as $E_{n k_z}(0) \to E_{n,-k_z}(0)$ and reverses all the winding numbers, $W_n(E) \to -W_n(E)$.
Accordingly, the boundary at which the skin modes localize and the surviving propagation direction are also reversed, whereas the CCW and CW characters of the magnon modes remain unchanged.
These relations are summarized in Table~\ref{tab:mirror}.
These results follow from the fact that the models with $\pm(D, D_\parallel)$ are mapped onto each other by the mirror reflection about the $xy$ plane, under which $k_z\to-k_z$ while the sense of rotation within the $xy$ plane is preserved.
For fixed signs of $D$ and $D_\parallel$, this mirror symmetry is broken, which allows the magnon dispersion to be asymmetric with respect to $k_z$ and the band minima to occur at nonzero momenta.
In fact, this symmetry breaking is necessary for the NHSE along the $z$ direction: If the system possessed a unitary symmetry mapping $k_z$ to $-k_z$, $\det(\tilde{H}_{k_z}-E)$ would be an even function of $k_z$, and the winding number would vanish for all $E$.
Therefore, the direction of the asymmetric propagation is determined by the chirality of the static texture through the sign of the band-minimum momentum $k_0$ in Eq.~(\ref{eq:winding_local_minimum}).
The rotational character of a magnon mode, by contrast, does not by itself determine the propagation direction: Being a CW mode is neither necessary nor sufficient for the band minimum to occur at a positive momentum.
Nevertheless, among the low-energy modes computed here, the CW mode is the most prone to the sign reversal of $k_0$ as the magnetic field is lowered.
This tendency can be understood from the mechanism identified in Ref.~\cite{sekiPropagationDynamicsSpin2020}: The DM-induced nonreciprocal part of the dispersion is governed by the overlap between the out-of-plane spin component of the static texture and the precession-amplitude distribution of each mode.
In our model, the precession amplitude of the CCW mode is concentrated in the outer region, whereas that of the CW mode is distributed over both the skyrmion core and the outer region, consistent with the distributions calculated in Ref.~\cite{sekiPropagationDynamicsSpin2020}.
Since $S_{\bm{r}}^{0,z}$ takes opposite signs at the core and in the outer region, the overlap remains sizable for the CCW mode, whereas it largely cancels for the CW mode.
The nonreciprocal part of the dispersion is therefore smaller for the CW mode than for the CCW mode, placing the CW band minimum closer to $k_z=0$ than the CCW band minimum.
Since the balance between the core and outer contributions depends on the magnetic field, the sign of the overlap, and hence of $k_0$, can reverse for the CW mode as the field is lowered.

\begin{table}
    \caption{\label{tab:mirror}Behavior of the relevant quantities under the sign reversal of the DM interaction, $\bm{D}_{\bm{r},\bm{r}'}\to-\bm{D}_{\bm{r},\bm{r}'}$, which is equivalent to the mirror reflection about the $xy$ plane.}
    \begin{ruledtabular}
        \begin{tabular}{ll}
            Quantity & $\bm{D}_{\bm{r},\bm{r}'}\to-\bm{D}_{\bm{r},\bm{r}'}$ \\
            \hline
            Chirality of the static skyrmion texture & Reversed \\
            Magnon dispersion $E_{nk_z}(0)$ & $E_{n,-k_z}(0)$ \\
            Band-minimum momentum $k_0$ & $-k_0$ \\
            Winding number $W_n(E)$ & $-W_n(E)$ \\
            Boundary hosting the skin modes & Top $\leftrightarrow$ bottom \\
            Surviving propagation direction & Reversed \\
            CCW/CW character of the magnon modes & Unchanged
        \end{tabular}
    \end{ruledtabular}
\end{table}

\section{\label{sec:conclusion}Conclusions and outlook}
We have theoretically studied the non-Hermitian point-gap topology of magnon bands in the presence of damping.
Using a linearized LLG equation with both local and nonlocal damping terms, we analytically evaluated the spectral winding number within perturbation theory, which predicts the occurrence of the NHSE.
Our analysis shows that the NHSE can occur even in the presence of only local damping, provided that the relative weights of the particle and hole components of magnons exhibit nontrivial $\bm{k}$ dependence.
Such momentum-dependent particle-hole mixing naturally arises in the magnon eigenmodes of noncollinear textures, whereas it is absent in collinear magnets with a single magnon band.
When nonlocal damping along the $z$ direction is included in the simplest nearest-neighbor model, we analytically show that the winding number of a given band is determined by the sign of the wave number at which the band attains its minimal energy.
These results are demonstrated employing a classical spin model hosting a skyrmion-string lattice as a steady state.
Furthermore, we investigate the spin-wave propagation dynamics excited by a pulsed magnetic field and show that the propagation direction drastically changes depending on the local and nonlocal damping terms and the excited bands, reflecting the underlying nontrivial winding numbers.

Here, we briefly discuss how our winding-number formula (\ref{eq:winding_formula}), obtained within the lowest-order nondegenerate perturbation theory, can be extended to energy bands with degeneracy points.
If unperturbed energy bands are well separated from each other, the higher-order corrections are expected to be negligible, and thus Eq.~(\ref{eq:winding_formula}) should remain valid.
Notably, even when the real unperturbed eigenenergies are degenerate, the degeneracy can be lifted in the imaginary parts of the complex-energy bands due to the difference in $\eta_{n\bm{k}}$.
In contrast, at degeneracy points, the complex-energy bands are generally not uniquely defined.
Moreover, if band hybridization occurs at higher orders in perturbation theory, it is generally impossible to assign the band indices to the unperturbed energy bands so that the exact complex-energy bands are smooth.
In such cases, the winding numbers should be calculated from the Hamiltonian projected onto the subspace spanned by the relevant bands.
Despite these difficulties, the hybridization is expected to be sufficiently small for skyrmion-string lattices, considered in this paper, provided that each band corresponds to distinct rotation and oscillation modes.
It would be interesting to extend the present formulation to systems with strong band hybridization or symmetry-protected degeneracies, where a more general treatment beyond nondegenerate perturbation theory may be required.
Furthermore, exploring the interplay between non-Hermitian topology and dynamical instabilities in such systems remains an important direction for future work.

While we have pointed out that the $\bm{k}$ dependence of $\eta_{n\bm{k}}$ plays an important role in the emergence of the NHSE, its physical origin remains unclear.
The mixing of the particle and hole components is necessary, but not all noncollinear spin configurations exhibit the nontrivial $\bm{k}$ dependence.
Further studies are needed to identify the precise conditions.
Another important direction is to propose an experimental setup that can directly demonstrate signatures of the NHSE in magnetic systems.
Periodic driving by lasers may be a possible way to manipulate damping parameters including nonlocal damping, as theoretically shown in Ref.~\cite{hanaiPhotoinducedNonreciprocalMagnetism2025}.

\begin{acknowledgments}
This work was supported by JSPS KAKENHI Grants No. JP24K00557, No. JP26H00385, and No. JP26K22289, as well as a Grant-in-Aid for JSPS Fellows (JSPS KAKENHI Grant No. JP24KJ1286).
\end{acknowledgments}

\appendix

\section{\label{sec:BdG_derivation}Derivation of Eq.~(\ref{eq:BdG})}
Here, we derive the non-Hermitian BdG equation~(\ref{eq:BdG}) by linearizing Eq.~(\ref{eq:generalized_LLG}) around a stationary solution $\bm{S}_{\bm{r}}^0$.
For a functional $f$ and column vectors $\bm{u}$ and $\bm{v}$, we define the second-derivative matrix by $[\delta^2 f/\delta\bm{u}\,\delta\bm{v}^\top]_{ij} \coloneqq \delta^2 f/\delta u_i\,\delta v_j$, i.e., the row (column) index corresponds to the variable without (with) the transpose.
The stationarity condition $\bm{S}_{\bm{r}}^0 \times \bm{B}_{\mathrm{eff},\bm{r}}^0 = \bm{0}$, where $\bm{B}_{\mathrm{eff},\bm{r}}^0 \coloneqq \bm{B}_{\mathrm{eff},\bm{r}}[\bm{S}_{\bm{r}}^0]$, implies that the effective field is parallel to the spin, $\bm{B}_{\mathrm{eff},\bm{r}}^0 = (\bm{B}_{\mathrm{eff},\bm{r}}^0 \cdot \tilde{\bm{e}}_{\bm{r}}^z)\, \tilde{\bm{e}}_{\bm{r}}^z$.
Substituting $\bm{S}_{\bm{r}} = \bm{S}_{\bm{r}}^0 + \delta\bm{S}_{\bm{r}}$ into Eq.~(\ref{eq:generalized_LLG}) and keeping the first order in $\delta\bm{S}_{\bm{r}}$, we obtain
\begin{align}
    \frac{\partial \delta\bm{S}_{\bm{r}}}{\partial t}
    &= \gamma \tilde{\bm{e}}_{\bm{r}}^z \times \left[S\, \delta\bm{B}_{\mathrm{eff},\bm{r}} - (\bm{B}_{\mathrm{eff},\bm{r}}^0 \cdot \tilde{\bm{e}}_{\bm{r}}^z)\, \delta\bm{S}_{\bm{r}}\right] \notag \\
    &\quad - \tilde{\bm{e}}_{\bm{r}}^z \times \sum_{\bm{r}'} \alpha_{\bm{r},\bm{r}'} \frac{\partial \delta\bm{S}_{\bm{r}'}}{\partial t}, \label{eq:linearized_LLG} \\
    \delta\bm{B}_{\mathrm{eff},\bm{r}} &= -\frac{1}{\hbar\gamma} \sum_{\bm{r}'} \frac{\delta^2 \mathcal{H}}{\delta \bm{S}_{\bm{r}} \delta \bm{S}_{\bm{r}'}^\top}[\bm{S}_{\bm{r}}^0]\, \delta\bm{S}_{\bm{r}'}.
\end{align}
Expanding Eq.~(\ref{eq:spin_wave_param}) to first order in $\psi_{\bm{r}}$ gives $\delta\bm{S}_{\bm{r}} = \sqrt{S} (\psi_{\bm{r}} \tilde{\bm{e}}_{\bm{r}}^- + \psi_{\bm{r}}^* \tilde{\bm{e}}_{\bm{r}}^+) + O(\psi^2)$.
Substituting this into Eq.~(\ref{eq:linearized_LLG}), projecting both sides onto the local transverse frame $(\tilde{\bm{e}}_{\bm{r}}^-, \tilde{\bm{e}}_{\bm{r}}^+)$, and using $\tilde{\bm{e}}_{\bm{r}}^\pm \times \tilde{\bm{e}}_{\bm{r}}^z = \pm\ii\, \tilde{\bm{e}}_{\bm{r}}^\pm$, we arrive at
\begin{align}
    \ii\hbar \sum_{\bm{r}'} \left(\tau_z \delta_{\bm{r},\bm{r}'} + \ii A_{\bm{r},\bm{r}'}\right) \frac{\partial}{\partial t}
    \begin{pmatrix}
        \psi_{\bm{r}'} \\
        \psi_{\bm{r}'}^*
    \end{pmatrix}
    &= \sum_{\bm{r}'} H_{\bm{r},\bm{r}'}
    \begin{pmatrix}
        \psi_{\bm{r}'} \\
        \psi_{\bm{r}'}^*
    \end{pmatrix}, \label{eq:linearized_projected}
\end{align}
where $A_{\bm{r},\bm{r}'}$ is defined in Eq.~(\ref{eq:A_def}) and
\begin{align}
    H_{\bm{r},\bm{r}'}
    &= S \begin{pmatrix}
        \tilde{\bm{e}}_{\bm{r}}^+{}^\top \\
        \tilde{\bm{e}}_{\bm{r}}^-{}^\top
    \end{pmatrix}
    \frac{\delta^2 \mathcal{H}}{\delta \bm{S}_{\bm{r}} \delta \bm{S}_{\bm{r}'}^\top}[\bm{S}_{\bm{r}}^0]
    \begin{pmatrix}
        \tilde{\bm{e}}_{\bm{r}'}^- & \tilde{\bm{e}}_{\bm{r}'}^+
    \end{pmatrix} \notag \\
    &\quad - \delta_{\bm{r},\bm{r}'} \left(\frac{\delta \mathcal{H}}{\delta \bm{S}_{\bm{r}}}[\bm{S}_{\bm{r}}^0] \cdot \tilde{\bm{e}}_{\bm{r}}^z\right) \tau_0, \label{eq:Hessian_spin}
\end{align}
with $\tau_0$ being the $2\times 2$ identity matrix acting on the Bogoliubov quasiparticle space.
Equation~(\ref{eq:Hessian_spin}) coincides with the Hessian matrix in Eq.~(\ref{eq:BdG_hamiltonian}), as shown below.
In terms of the Nambu spinor $\bm{\Psi}_{\bm{r}} \coloneqq (\psi_{\bm{r}}, \psi_{\bm{r}}^*)^\top$, the Hessian matrix in Eq.~(\ref{eq:BdG_hamiltonian}) reads $\delta^2\mathcal{H}/\delta\bm{\Psi}_{\bm{r}}^* \delta\bm{\Psi}_{\bm{r}'}^\top$ evaluated at $\psi_{\bm{r}} = 0$.
The chain rule for the composition $\mathcal{H}[\bm{S}_{\bm{r}}(\bm{\Psi}_{\bm{r}})]$ given by Eq.~(\ref{eq:spin_wave_param}) yields
\begin{align}
    \frac{\delta^2 \mathcal{H}}{\delta\bm{\Psi}_{\bm{r}}^* \delta\bm{\Psi}_{\bm{r}'}^\top}
    &= \frac{\delta \bm{S}_{\bm{r}}^\top}{\delta\bm{\Psi}_{\bm{r}}^*}
    \frac{\delta^2 \mathcal{H}}{\delta \bm{S}_{\bm{r}} \delta \bm{S}_{\bm{r}'}^\top}
    \frac{\delta \bm{S}_{\bm{r}'}}{\delta\bm{\Psi}_{\bm{r}'}^\top}
    + \delta_{\bm{r},\bm{r}'} \frac{\delta \mathcal{H}}{\delta \bm{S}_{\bm{r}}} \cdot \frac{\delta^2 \bm{S}_{\bm{r}}}{\delta\bm{\Psi}_{\bm{r}}^* \delta\bm{\Psi}_{\bm{r}}^\top},
    \label{eq:chain_rule}
\end{align}
where the second term is proportional to $\delta_{\bm{r},\bm{r}'}$ because $\bm{S}_{\bm{r}}$ depends only on $\bm{\Psi}_{\bm{r}}$, and the dot denotes the contraction over the spin components, $(\delta \mathcal{H}/\delta \bm{S}_{\bm{r}}) \cdot (\delta^2 \bm{S}_{\bm{r}}/\delta\bm{\Psi}_{\bm{r}}^* \delta\bm{\Psi}_{\bm{r}}^\top) = \sum_{\mu} (\delta \mathcal{H}/\delta S_{\bm{r}}^\mu)(\delta^2 S_{\bm{r}}^\mu/\delta\bm{\Psi}_{\bm{r}}^* \delta\bm{\Psi}_{\bm{r}}^\top)$.
Setting $\psi_{\bm{r}} = 0$ in Eq.~(\ref{eq:spin_wave_param}) gives $\delta\bm{S}_{\bm{r}}/\delta\bm{\Psi}_{\bm{r}}^\top = \sqrt{S} \begin{pmatrix}
    \tilde{\bm{e}}_{\bm{r}}^- & \tilde{\bm{e}}_{\bm{r}}^+
\end{pmatrix}$ and $\delta^2\bm{S}_{\bm{r}}/\delta\bm{\Psi}_{\bm{r}}^* \delta\bm{\Psi}_{\bm{r}}^\top = -\tilde{\bm{e}}_{\bm{r}}^z \tau_0$, while the functional derivatives of $\mathcal{H}$ are evaluated at $\bm{S}_{\bm{r}}^0$.
The first and second terms of Eq.~(\ref{eq:chain_rule}) then reproduce those of Eq.~(\ref{eq:Hessian_spin}), respectively.
Multiplying Eq.~(\ref{eq:linearized_projected}) by $(\tau_z + \ii A)^{-1}$ from the left then yields Eqs.~(\ref{eq:BdG}) and (\ref{eq:magnon_nh_hamiltonian}).

We also note that the Hessian matrix~(\ref{eq:Hessian_spin}) is identical to the coefficient matrix obtained by expanding the corresponding quantum spin Hamiltonian up to quadratic order in the magnon creation and annihilation operators $\hat{\bm{a}}^\dagger = (\hat{a}_{\bm{r}}^\dagger)$ and $\hat{\bm{a}} = (\hat{a}_{\bm{r}})$ introduced by the Holstein-Primakoff transformation~\cite{holsteinFieldDependenceIntrinsic1940}:
\begin{align}
    \hat{\mathcal{H}} &= E_0 + \frac{1}{2}
    \begin{pmatrix}
        \hat{\bm{a}}^\dagger & \hat{\bm{a}}
    \end{pmatrix}
    H
    \begin{pmatrix}
        \hat{\bm{a}} \\
        \hat{\bm{a}}^\dagger
    \end{pmatrix}
    + O(\hat{a}^3), \label{eq:HP_expansion}
\end{align}
where $E_0 = \mathcal{H}[\bm{S}_{\bm{r}}^0]$ is the energy of the stationary state.
We emphasize, however, that our analysis is based on classical spins rather than on the quantum-mechanical formulation in terms of $\hat{\bm{a}}$ and $\hat{\bm{a}}^\dagger$.
This is because the Gilbert damping is formulated for classical spins through the LLG equation~(\ref{eq:generalized_LLG}), and hence its effect on magnons can be incorporated precisely only within calculations based on classical spins.

\section{\label{sec:left_derivation}Derivation of Eq.~(\ref{eq:left_right_relation})}
Here, we derive Eq.~(\ref{eq:left_right_relation}), which relates the left eigenstate $\ket{\phi_n^\mathrm{L}(0)}$ to the right eigenstate $\ket{\phi_{n^*}^\mathrm{R}(0)}$ and allows us to eliminate $\ket{\phi_n^\mathrm{L}(0)}$ from Eq.~(\ref{eq:eff_damping}).
Multiplying the eigenvalue equation $\tilde{H}(0) \ket{\phi_{n^*}^\mathrm{R}(0)} = E_{n^*}(0) \ket{\phi_{n^*}^\mathrm{R}(0)}$ by $\tau_z$ from the left and using the pseudo-Hermiticity $\tau_z \tilde{H}(0) \tau_z^{-1} = \tilde{H}^\dagger(0)$, we obtain
\begin{align}
    \tilde{H}^\dagger(0)\, \tau_z \ket{\phi_{n^*}^\mathrm{R}(0)} &= E_n^*(0)\, \tau_z \ket{\phi_{n^*}^\mathrm{R}(0)},
\end{align}
where we have used $E_{n^*}(0) = E_n^*(0)$.
Thus, $\tau_z \ket{\phi_{n^*}^\mathrm{R}(0)}$ is an eigenstate of $\tilde{H}^\dagger(0)$ with the eigenvalue $E_n^*(0)$, which is the defining property of the left eigenstate $\ket{\phi_n^\mathrm{L}(0)}$.
Since the eigenvalue $E_n(0)$ of $\tilde{H}(0)$ is assumed to be nondegenerate, the eigenvalue $E_n^*(0)$ of $\tilde{H}^\dagger(0)$ is also nondegenerate, and hence $\tau_z \ket{\phi_{n^*}^\mathrm{R}(0)}$ is proportional to $\ket{\phi_n^\mathrm{L}(0)}$.
Taking the inner product with $\ket{\phi_n^\mathrm{R}(0)}$ and using the normalization $\braket{\phi_n^\mathrm{L}(0)|\phi_n^\mathrm{R}(0)}=1$, we fix the proportionality constant and arrive at Eq.~(\ref{eq:left_right_relation}).

\section{\label{sec:winding_derivation}Derivation of Eq.~(\ref{eq:winding_formula})}
From the definition of the winding number in Eq.~(\ref{eq:spectral_winding_number}) and the perturbative result given in Eq.~(\ref{eq:perturbed_energy_k}), we show that there exists a complex energy $E$ satisfying Eq.~(\ref{eq:winding_formula}).
We use the generalized Nielsen-Ninomiya theorem~\cite{besshoNielsenNinomiyaTheoremBulk2021}, which relates the winding number for a single band to the sign of its group velocity:
\begin{align}
    -W_n(E) &= \sum_{\substack{\Real E_{n k_\parallel} = \Real E \\ \Imag E_{n k_\parallel} > \Imag E}} \sgn v_{n k_\parallel}, \label{eq:nn_theorem}
\end{align}
where $v_{n k_\parallel}=\Real \partial_{k_\parallel} E_{n k_\parallel}/\hbar$ is the group velocity for the band with complex eigenenergy $E_{n k_\parallel}$ and the summation is taken over $k_\parallel$ satisfying $\Real E_{n k_\parallel} = \Real E$ and $\Imag E_{n k_\parallel} > \Imag E$.

Before presenting the detailed derivation, we briefly outline the strategy of the proof.
For simplicity, we assume that $\Real E_{n k_\parallel}$ has a single (and hence global) minimum at $k_\parallel=k_0$.
Under this assumption, the Taylor expansion of the band around $k_\parallel=k_0$ yields $\sgn v_{n, k_0 \pm \Delta k}=\pm 1$, where $\Delta k$ is a positive infinitesimal momentum.
Thus, from Eq.~(\ref{eq:nn_theorem}), there exists a complex energy $E$ with $W_n(E)\neq 0$ only if $\Imag E_{n k_\parallel}$ takes different values at $k_\parallel=k_0\pm \Delta k$.
In other words, only if the lifetimes at $k_\parallel=k_0 \pm \Delta k$ differ does the one-way propagating eigenmode with longer lifetime persist in the long-time limit, thereby forming the skin modes with eigenenergy $E$ satisfying $W_n(E)\neq 0$ under the $x_\parallel$-OBC.
To summarize, the essential information for determining the winding number $W_n(E)$ is the position of the minimum point $k_\parallel=k_0$ of $\Real E_{n k_\parallel}$ and the relative magnitude of the lifetimes ($\Imag E_{n k_\parallel}$) at $k_\parallel=k_0 \pm \Delta k$.

We now turn to the detailed derivation.
We assume that the $n$th magnon band energy $E_{n k_\parallel}(0)$ in the absence of damping is real and has a unique minimum $E_0$ at $k_\parallel=k_0$.
We also assume that $E_{n k_\parallel}(0)$ is nondegenerate in the vicinity of its minimum point $k_\parallel=k_0$.
Then, we can use Eq.~(\ref{eq:perturbed_energy_k}) around the minimum point, whose real part $\Real E_{n k_\parallel}(\alpha) = E_{n k_\parallel}(0)$ can be expanded around $k_\parallel=k_0$ as
\begin{align}
    E_{n k_\parallel}(0) &= E_0 + \frac{\hbar^2}{2m_*} (k_\parallel - k_0)^2 + O\left((k_\parallel - k_0)^3\right),
\end{align}
where $E_0\in\mathbb{R}$ is a band minimum and $m_*>0$ is the effective mass at $k_\parallel=k_0$ defined by
\begin{align}
    m_* \coloneqq \left.\left(\frac{1}{\hbar^2} \frac{\partial^2 E_{n k_\parallel}(0)}{\partial k_\parallel^2}\right)^{-1}\right|_{k_\parallel=k_0}.
\end{align}
Defining $\Real E \coloneqq E_0 + \frac{\hbar^2}{2m_*} \Delta k^2$ with an infinitesimal $\Delta k>0$, we obtain the momenta satisfying $\Real E_{n k_\parallel}(\alpha)=E_{n k_\parallel}(0)=\Real E$ being
\begin{align}
    k_\parallel &= k_0 \pm \Delta k + O(\Delta k^2).
\end{align}
At these momenta the group velocity becomes
\begin{align}
    v_{n, k_0 \pm \Delta k} &= \pm \frac{\hbar}{m_*} \Delta k + O(\Delta k^2),
\end{align}
leading to
\begin{align}
    \sgn v_{n, k_0 \pm \Delta k} &= \pm 1.
\end{align}

In the following, we examine the condition under which the imaginary part $\Imag E_{n k_\parallel}(\alpha)=-E_{n k_\parallel}(0) \alpha_{\mathrm{eff},n k_\parallel}$ of the eigenenergy differs between these momenta.
If they differ, $W(E)\neq 0$ for $\Imag E$ between these values; otherwise, $W(E)=0$.
With this in mind, the above condition can be reformulated as the condition under which the following function does not vanish:
\begin{align}
    f_n(k_0) &\coloneqq \Imag E_{n, k_0 + \Delta k}(\alpha) - \Imag E_{n, k_0 - \Delta k}(\alpha) \notag \\
    &= -\Real E\, (\alpha_{\mathrm{eff},n, k_0 + \Delta k} - \alpha_{\mathrm{eff},n, k_0 - \Delta k}).
\end{align}
To determine whether $f_n(k_0)$ is nonzero, it suffices to evaluate its sign.
We perform the Taylor expansion of $\alpha_{\mathrm{eff},n k_\parallel}$ around $k_\parallel=k_0$, which yields
\begin{align}
    f_n(k_0) &= -2\Real E \sum_{l=0}^{\infty} \frac{\Delta k^{2l + 1}}{(2l + 1)!} \left.\partial_{k_\parallel}^{2l + 1} \alpha_{\mathrm{eff},n k_\parallel}\right|_{k_\parallel=k_0}.
\end{align}
Since $\sgn\Real E=\sgn E_0$ for $E_0\neq 0$ and sufficiently small $\Delta k$, whereas $\sgn\Real E=+1$ for $E_0=0$, we shall use the notation $\sgn E_0=+1$ for $E_0=0$.
Then we have
\begin{align}
    \sgn f_n(k_0) &= -\sgn E_0 \sgn\left.\partial_{k_\parallel}^{2l_0 + 1} \alpha_{\mathrm{eff},n k_\parallel}\right|_{k_\parallel=k_0},
\end{align}
where $l_0$ is the smallest $l=0,1,\dots$ such that $\left.\partial_{k_\parallel}^{2l_0 + 1} \alpha_{\mathrm{eff},n k_\parallel}\right|_{k_\parallel=k_0} \neq 0$.
If $\left.\partial_{k_\parallel}^{2l_0 + 1} \alpha_{\mathrm{eff},n k_\parallel}\right|_{k_\parallel=k_0} = 0$ for all $l=0,1,\dots$, then $f_n(k_0) = 0$.
For simplicity, we consider only the case $\left.\partial_{k_\parallel} \alpha_{\mathrm{eff},n k_\parallel}\right|_{k_\parallel=k_0} \neq 0$, i.e., $l_0=0$.
Then, we can always find $E$ satisfying Eq.~(\ref{eq:winding_formula}) by choosing $\Imag E$ between $\Imag E_{n, k_0 - \Delta k}(\alpha)$ and $\Imag E_{n, k_0 + \Delta k}(\alpha)$.
This result is particularly intuitive.
Let us consider a particle band with $E_0>0$.
If $\alpha_{\mathrm{eff},n k_\parallel} \geq 0$ increases (decreases) monotonically around $k_\parallel=k_0$, the damping strength at $k_\parallel=k_0 - \Delta k$ ($k_\parallel=k_0 + \Delta k$) is smaller than that at the other momentum.
Consequently, the eigenmode at $k_\parallel=k_0-\Delta k$ ($k_\parallel=k_0+\Delta k$) dominates in the long-time limit.

Finally, we comment on the possibility of $|W_n(E)|>1$, which is avoided in the derivation of Eq.~(\ref{eq:winding_formula}) by assuming the unique global band-minimum.
This situation may occur under a suitable choice of $E$ when $E_{n k_\parallel}(0)$ has two or more local minima.
We remember that the real part of $E$ has been chosen to be slightly larger than the global minimum $E_0$.
If we take $E$ as $\Real E>E_1$, where $E_1$ is a local but not global minimum, the condition of the summation in Eq.~(\ref{eq:nn_theorem}), $\Real E_{n k_\parallel}(\alpha) = E_{n k_\parallel}(0) = \Real E$, is satisfied at four or more momenta.
In this case, $W_n(E)$ is determined from the summation of Eq.~(\ref{eq:winding_formula}) for these momenta, and thus $|W_n(E)|$ can exceed 1.

\bibliography{reference}

\end{document}